\documentclass[twocolumn,aps,showpacs,prb,tightenlines,amsmath,amssymb]{revtex4}
\usepackage{graphicx}
\usepackage{amssymb}
\usepackage{amsmath}
\usepackage{colordvi}

\begin{document}

\title{Electron spin relaxation in bulk III-V semiconductors from
 a fully microscopic kinetic spin Bloch equation approach} 
\author{J. H. Jiang}
\author{M. W. Wu}
\thanks{Author to whom correspondence should be addressed}
\email{mwwu@ustc.edu.cn.}
\affiliation{Hefei National Laboratory for Physical Sciences at
  Microscale, University of Science and Technology of China, Hefei,
  Anhui, 230026, China}
\affiliation{Department of Physics,
  University of Science and Technology of China, Hefei,
  Anhui, 230026, China}
\altaffiliation{Mailing Address}

\date{\today}

\begin{abstract} 
Electron spin relaxation in bulk III-V semiconductors is investigated
from a fully microscopic kinetic spin Bloch equation approach where all
relevant scatterings, such as, the electron--nonmagnetic-impurity,
electron-phonon, electron-electron, electron-hole, and
electron-hole exchange (the Bir-Aronov-Pikus mechanism) scatterings are
explicitly included. The Elliot-Yafet mechanism is also fully
incorporated. This approach offers a way toward thorough understanding
of electron spin relaxation both near and far away from the equilibrium
in the metallic regime. The dependences of the spin
relaxation time on electron density, temperature, initial spin polarization,
photo-excitation density, and hole density are studied
thoroughly with the underlying physics analyzed. We find that these
dependences are usually {\em qualitatively} different in the
non-degenerate and degenerate regimes. In contrast to the 
previous investigations in the literature, we find that: (i) In
$n$-type materials, the Elliot-Yafet mechanism is {\em less} important than
the D'yakonov-Perel' mechanism, even for the narrow band-gap
semiconductors such as InSb and InAs. (ii) The 
density dependence of the spin relaxation time is nonmonotonic and we
predict a {\em peak} in the metallic regime in both $n$-type and intrinsic
materials. (iii) In intrinsic materials, the Bir-Aronov-Pikus mechanism is
found to be negligible compared with the D'yakonov-Perel'
mechanism. We also predict a peak in the temperature dependence
of spin relaxation time which is due to the nonmonotonic temperature
dependence of the electron-electron Coulomb scattering in intrinsic
materials with small initial spin polarization. (iv) In
$p$-type III-V semiconductors, the Bir-Aronov-Pikus mechanism 
dominates spin relaxation in the low temperature regime only when the
photo-excitation density is low. When the photo-excitation
density is high, the Bir-Aronov-Pikus mechanism can be
comparable with the D'yakonov-Perel' mechanism only in the
moderate temperature regime roughly around the Fermi temperature
of electrons, whereas for higher or lower temperature it is
unimportant. The relative importance of the Bir-Aronov-Pikus mechanism
decreases with the photo-excitation density and eventually
becomes negligible at sufficiently high photo-excitation density.
The effect of electric field on spin relaxation in $n$-type III-V
semiconductors is also studied with behaviors very different from
those in the two-dimensional case reported. Finally, we find good
agreement of our calculation with the experimental results.
\end{abstract}


\maketitle

\section{Introduction}

Semiconductor spintronics, which aims at utilizing or incorporating
the spin degree of freedom in electronics, has attracted much
interest.\cite{Wolf,spintronics,opt-or} During the last decade, the 
fast developments of techniques of coherent manipulation of 
electron spins via optical or
electrical methods\cite{opt-nmr,Oestreich,spin-coherent0,spin-coherent1,Kato1,Kato,spin-coherent2,Awsch98,longSRT,Awschtrans}
have intrigued a lot of studies.\cite{spintronics} Many of
the important findings are developed in bulk GaAs or GaAs epilayers,
where the spin relaxation time (SRT) was found to be as long as
130~ns,\cite{Awsch98,longSRT} and the spin diffusion length was reported as large as
100~$\mu$m.\cite{Awschtrans} Remarkably, it has been found that the
SRT can vary by more than three orders of magnitude
with temperature or electron density.\cite{Awsch98} 
The relevant spin relaxation mechanisms for electron system in the
metallic regime have been recognized for a long time as: (i)
the Elliot-Yafet (EY) mechanism in which electron spins have a small
chance to flip during each scattering due to spin
mixing in the conduction band;\cite{EY} (ii) the D'yakonov-Perel' (DP)
mechanism in which the electron spins decay due to their precession
around the {\bf k}-dependent spin-orbit fields (inhomogeneous broadening\cite{wu-early}) during the free flight
between adjacent scattering events;\cite{DP} (iii) the
Bir-Aronov-Pikus (BAP) mechanism in which electrons exchange their
spins with holes.\cite{BAP} The hyperfine interaction is another
mechanism which is usually important for spin relaxation of localized
electrons and ineffective in metallic regime where most of the carriers
are in extended states.\cite{Pershin,localize,localize2,opt-nmr,opt-or,spintronics,localizeReview}

Despite decades of study, a detailed theoretical investigation from
a fully microscopic approach in bulk system has not yet been
performed. Although Song and Kim have investigated electron spin
relaxation due to all the three relevant mechanisms for various
conditions in both $n$- and $p$-type semiconductors,\cite{Song} in
their work they use the analytical expressions based on single-particle
approach which are only
applicable for the non-degenerate electron system, and hence make the
discussion in low temperature and/or high density regime
questionable. More importantly, the carrier-carrier Coulomb
scattering, which has been shown to be very important for spin
relaxation in two-dimensional
systems,\cite{wu-early,wu-review,Harley,Ivchenko,Ji,lowT,highP,hot-e,multi-band,multi-valley,wu-hole,wu-bap,terahertz,Lai} 
has not yet been well studied in bulk system. Also, at finite spin
polarization, the Coulomb Hartree-Fock (HF) term acts as an effective
longitudinal magnetic field, which has been demonstrated to be
able to increase the SRT by more than one order of magnitude when the
initial spin polarization is high in two-dimensional electron system (2DES).\cite{highP,wu-exp-hP,tobias} However, the effect of the
Coulomb HF term in bulk system has not been investigated.
Another issue is that the commonly used analytical formula for spin
relaxation due to the BAP mechanism is based on the elastic scattering
approximation, which has been proved to be invalid for low
temperature due to pretermission of the Pauli blocking of electrons very recently by
Zhou and Wu.\cite{wu-bap} 
Consequently, the BAP mechanism has been demonstrated to be
unimportant in two-dimensional system at moderate and high excitation density first
theoretically\cite{wu-bap} and then experimentally,\cite{Cui} which is 
in stark contrast with the common belief in the literature. Whether it is
still true in bulk system remains unchecked.
Furthermore, in most works only the short-range 
electron-hole exchange interaction was
considered, while the long-range part was
ignored.\cite{opt-or,Song,Fishman,Zerrouati,spintronics}
All these questions suggest that a detailed fully microscopic investigation is
needed. In this work, we perform such a study from the fully
microscopic kinetic spin Bloch equation (KSBE)
approach.\cite{wu-early,highP,hot-e,lowT,nonMarkov,wu-bap,terahertz,multi-band,multi-valley,wu-review,wu-hole}
We focus on the metallic regime where most of the carriers
are in extended states. We restrict ourselves to the zero magnetic field
case.\cite{magB}

Previously, the KSBE approach has been applied extensively to study spin
dynamics in semiconductor nanostructures in both Markovian and
non-Markovian limits and in systems both near and far away from the
equilibrium (under strong static or terahertz electric field or with high spin
polarization).\cite{wu-early,wu-strain,wu-hole,highP,hot-e,lowT,nonMarkov,wu-bap,terahertz,wu-exp,wu-exp-hP,wu-review,multi-valley,multi-band,Lai} 
The KSBE approach has been demonstrated to be successful in the study
of spin relaxation in semiconductor quantum wells where good agreements
with experiments have been achieved and many predictions have been confirmed by 
experiments.\cite{wu-exp,Awsch-exp,wu-exp-hP,multi-band,Ji,Zheng,Zheng2,Lai,tobias,Cui} 
This approach has also been applied in the investigation of spin
relaxation/dephasing in bulk GaAs many years ago by Wu and Ning,
where only the electron--nonmagnetic-impurity and
electron-phonon scatterings are included.\cite{wu-3d}
In this work, we include all the scatterings, especially the
electron-electron Coulomb, electron-hole Coulomb, and electron-hole
exchange scatterings together with the EY mechanism which were not
considered in Ref.~\onlinecite{wu-3d}.

An important goal of this work is to find the dominant mechanism in
different parameter regimes and for different materials from the fully
microscopic KSBE approach. Previous investigations in the literature indicate that for
$n$-type III-V semiconductors, the spin relaxation is mostly dominated by
the DP mechanism, except at low temperature where the EY mechanism is most
important; for intrinsic and $p$-type III-V semiconductors, the BAP mechanism
dominates at low temperature when the hole density is high, 
whereas the DP mechanism dominates in other
regimes.\cite{Aronov,spintronics,Song,opt-or} In the present
work, from the fully microscopic KSBE approach, however, we find that 
the EY mechanism is less important than the DP mechanism in
$n$-type III-V semiconductors even for narrow band-gap
semiconductors, such as InAs and InSb. For $p$-type III-V
semiconductors, we find that the BAP mechanism 
dominates spin relaxation in the low temperature regime only when the
photo-excitation density is low enough. However, when the photo-excitation
density is high, the BAP mechanism can be comparable
with the DP mechanism only in the moderate temperature regime
roughly around the Fermi temperature of electrons, and for higher or
lower temperature it is unimportant. The relative importance of the
BAP mechanism decreases with the photo-excitation density and
eventually becomes negligible for sufficiently high photo-excitation
density. For intrinsic III-V semiconductors, the
BAP mechanism is negligible.

An important method of spin injection is the hot-electron spin
injection where high electric field is applied.\cite{biasInjec} 
Moreover, in 2DES, the spin relaxation can be
effectively manipulated by the high in-plane electric
field.\cite{hot-e,multi-band,multi-valley} In
this work, we also study spin relaxation in $n$-type bulk
semiconductors under high electric field. We show that there is some
essential difference of the electric field effect on spin dynamics
between 2DES and the bulk system. Using GaAs as an example, we
demonstrate that the electric field dependence of spin lifetime can be
nonmonotonic or monotonic depending on the lattice temperature and the densities of
impurities and electrons. The underlying physics is analyzed. The
results indicate that the spin lifetime can be effectively controlled by
electric field.

This paper is organized as follows: In Sec.~II, we introduce the
KSBEs. In Sec.~III, we study the spin relaxation in $n$-type
III-V semiconductors. In Secs.~IV and V, we investigate the spin relaxations in
intrinsic and $p$-type III-V semiconductors, respectively. We study the
effects of electric field on spin relaxation in $n$-type III-V
semiconductors in Sec.~VI. Finally, we conclude in Sec.~VII.

\section{KSBEs}

The spin dynamics is studied by solving the microscopic KSBEs derived
via the nonequilibrium Green function method,\cite{Haug,wu-early,highP,Franz} 
\begin{equation}
  \label{KSBE}
  \partial_{t}\hat{\rho}_{{\bf k}} = 
  \partial_{t}\hat{\rho}_{{\bf k}}|_{\rm coh} +
  \partial_{t}\hat{\rho}_{{\bf k}}|_{\rm drift} +
  \partial_{t}\hat{\rho}_{{\bf k}}|_{\rm scat}.
\end{equation}
Here $\hat{\rho}_{\bf k}$ is the single particle density matrix with
the diagonal terms representing the distributions of each spin band, and
the off-diagonal terms denoting the correlation of the two spin bands.
The coherent term is given by
\begin{equation}
  \partial_{t}\hat{\rho}_{{\bf k}}|_{\rm coh} = - i[{\bf \Omega}({\bf k})
    \cdot \frac{\mbox{\boldmath$\hat{\sigma}$\unboldmath}}{2} +
    \hat{\Sigma}_{\rm HF}({\bf k}), \hat{\rho}_{{\bf k}}].
\end{equation}
Here $[\ ,\ ]$ is the commutator and ${\bf \Omega}({\bf k}) = {\bf
  \Omega}_{\rm D}({\bf k}) + {\bf \Omega}_{\rm S}({\bf k})$ with
\begin{equation}
  {\bf \Omega}_{\rm D}({\bf k}) = 2\gamma_{\rm D} \left(k_x\left(k_y^2-k_z^2\right),
  k_y\left(k_z^2-k_x^2\right), k_z\left(k_x^2-k_y^2\right)\right)
\label{Dresselsoc}
\end{equation}
due to the Dresselhaus spin-orbit coupling (SOC)\cite{Dressel} and
\begin{equation}
  {\bf \Omega}_{\rm S}({\bf k}) = 2 \beta \left(k_x, -k_y, 0\right)
\end{equation}
due to the strain induced SOC.\cite{opt-or,Kato,wu-strain}
$\hat{\Sigma}_{\rm HF}({\bf k})=-\sum_{{\bf k}^{\prime}}V_{{\bf k}-{\bf
k}^{\prime}}\hat{\rho}_{{\bf k}^{\prime}}$ is the Coulomb HF
term of the electron-electron interaction. Previously, it
was found that the Coulomb HF term serves as a longitudinal effective
magnetic field which increases with the initial spin
polarization. The effective magnetic field can be as large
as 40~T, which blocks the inhomogeneous broadening of the ${\bf
  k}$-dependent spin-orbit field and reduces the spin relaxation due
to the DP mechanism in 2DES.\cite{highP}
However, the effect of the HF effective magnetic field on
spin relaxation in bulk system is still unknown, which is one of
the goal of this paper.
The drift term is given by,\cite{hot-e}
\begin{equation}
\partial_{t}\hat{\rho}_{{\bf k}}|_{\rm drift} = - e {\bf
  E}\cdot\mbox{\boldmath$\nabla$\unboldmath}_{\bf k} \hat{\rho}_{{\bf
  k}}, 
\end{equation}
where $e$ is the electron charge ($e<0$) and ${\bf E}$ is the electric field. 
As the hole spin relaxation is very fast ($\sim$100~fs),\cite{Hilton,Schneider} one can assume that the hole
system is kept in the thermal equilibrium state where the hole
distribution ($f^{h}_{{\bf k},m}$) is described by the Fermi
distribution. The scattering term
$\partial_{t}\hat{\rho}_{{\bf k}}|_{\rm scat}$ contains the contributions
from the electron-impurity scattering $\partial_{t}\hat{\rho}_{{\bf
    k}}|_{\rm ei}$, the electron-phonon scattering
$\partial_{t}\hat{\rho}_{{\bf k}}|_{\rm ep}$, the electron-electron
scattering $\partial_{t}\hat{\rho}_{{\bf k}}|_{\rm ee}$, the electron-hole Coulomb scattering
$\partial_{t}\hat{\rho}_{{\bf k}}|_{\rm eh}$, and the electron-hole exchange scattering
$\partial_{t}\hat{\rho}_{{\bf k}}|_{\rm ex}$,
\begin{eqnarray}
\partial_{t}\hat{\rho}_{{\bf k}}|_{\rm scat} &=&
\partial_{t}\hat{\rho}_{{\bf k}}|_{\rm ei} +
\partial_{t}\hat{\rho}_{{\bf k}}|_{\rm ep} +
\partial_{t}\hat{\rho}_{{\bf k}}|_{\rm ee} \nonumber \\ &&\mbox{} +
\partial_{t}\hat{\rho}_{{\bf k}}|_{\rm eh} +
\partial_{t}\hat{\rho}_{{\bf k}}|_{\rm ex}.
\end{eqnarray}
These terms read
\begin{eqnarray}
\left. \partial_{t}\hat{\rho}_{{\bf k}}\right|_{\rm ei} &=& - \pi \sum_{{\bf
      k}^{\prime}} n_i Z_i^2 V_{{\bf k}-{\bf k}^{\prime}}^2
  \delta(\varepsilon_{{\bf k}^{\prime}}-\varepsilon_{{\bf k}})
      \Big(
    \hat{{\Lambda}}_{{\bf k},{\bf k}^{\prime}} \hat{\rho}^{>}_{{\bf
  k}^{\prime}}\hat{{\Lambda}}_{{\bf k}^{\prime},{\bf k}} \nonumber \\
      &&\mbox{} \times \hat{\rho}^{<}_{{\bf 
  k}} - \hat{{\Lambda}}_{{\bf k},{\bf k}^{\prime}} \hat{\rho}^{<}_{{\bf
  k}^{\prime}}\hat{{\Lambda}}_{{\bf k}^{\prime},{\bf k}} \hat{\rho}^{>}_{{\bf
  k}} \Big)  + {\rm H.c.},
\label{scat_ei}
\end{eqnarray}
\begin{eqnarray}
\left. \partial_{t}\hat{\rho}_{{\bf k}}\right|_{\rm ep} &=& - \pi \sum_{\lambda,\pm,{\bf
  k}^{\prime}} |M_{\lambda,{\bf k}-{\bf k}^{\prime}}|^2
  \delta(\pm\omega_{\lambda,{\bf k}-{\bf
  k}^{\prime}}+\varepsilon_{{\bf k}^{\prime}}-\varepsilon_{{\bf k}})
  \nonumber \\ &&\mbox{} \times
  \Big(N_{\lambda,{\bf k}-{\bf k}^{\prime}}^{\pm} \hat{{\Lambda}}_{{\bf
  k},{\bf k}^{\prime}} \hat{\rho}^{>}_{{\bf k}^{\prime}}\hat{{\Lambda}}_{{\bf
  k}^{\prime},{\bf k}} \hat{\rho}^{<}_{{\bf k}} - 
    N_{\lambda,{\bf k}-{\bf k}^{\prime}}^{\mp} \hat{{\Lambda}}_{{\bf k},{\bf
    k}^{\prime}} 
\nonumber \\ &&\mbox{} \times 
\hat{\rho}^{<}_{{\bf k}^{\prime}}  
\hat{{\Lambda}}_{{\bf k}^{\prime},{\bf k}} 
\hat{\rho}^{>}_{{\bf k}} \Big) + {\rm H.c.} , 
\label{scat_ep}
\end{eqnarray}
\begin{widetext}
\begin{eqnarray}
\left. \partial_{t}\hat{\rho}_{{\bf k}}\right|_{\rm ee} &=& - \pi \sum_{{\bf
      k}^{\prime},{\bf
      k}^{\prime\prime}} V_{{\bf k}-{\bf k}^{\prime}}^2
  \delta(\varepsilon_{{\bf k}^{\prime}}-\varepsilon_{{\bf
      k}}+\varepsilon_{{\bf k}^{\prime\prime}}-\varepsilon_{{\bf
      k}^{\prime\prime}-{\bf k}+{\bf k}^{\prime}}) 
  \Big[ \hat{{\Lambda}}_{{\bf k},{\bf k}^{\prime}} \hat{\rho}^{>}_{{\bf
      k}^{\prime}} \hat{{\Lambda}}_{{\bf k}^{\prime},{\bf k}}
      \hat{\rho}^{<}_{{\bf k}} {\rm Tr}\left(\hat{{\Lambda}}_{{\bf
      k}^{\prime\prime},{\bf k}^{\prime\prime}-{\bf k}+{\bf
      k}^{\prime}} \hat{\rho}^{<}_{{\bf k}^{\prime\prime}-{\bf k}+{\bf k}^{\prime}} 
      \hat{{\Lambda}}_{{\bf k}^{\prime\prime}-{\bf k}+{\bf k}^{\prime},{\bf
	  k}^{\prime\prime}} \hat{\rho}^{>}_{{\bf
      k}^{\prime\prime}}\right) \nonumber\\ &&
      \mbox{\ \ } - \hat{{\Lambda}}_{{\bf k},{\bf k}^{\prime}}
      \hat{\rho}^{<}_{{\bf 
      k}^{\prime}} \hat{{\Lambda}}_{{\bf k}^{\prime},{\bf k}}
      \hat{\rho}^{>}_{{\bf k}} {\rm Tr}\left(\hat{{\Lambda}}_{{\bf
      k}^{\prime\prime},{\bf k}^{\prime\prime}-{\bf k}+{\bf
      k}^{\prime}} \hat{\rho}^{>}_{{\bf k}^{\prime\prime}-{\bf k}+{\bf k}^{\prime}} 
      \hat{{\Lambda}}_{{\bf k}^{\prime\prime}-{\bf k}+{\bf k}^{\prime},{\bf
	  k}^{\prime\prime}} \hat{\rho}^{<}_{{\bf
      k}^{\prime\prime}}\right) \Big]+ {\rm H.c.}, 
\label{scat_ee}
\end{eqnarray}
\begin{eqnarray}
\left. \partial_{t}\hat{\rho}_{{\bf k}}\right|_{\rm eh} &=& - \pi \sum_{{\bf
      k}^{\prime},{\bf
    k}^{\prime\prime},m,m^{\prime}} V_{{\bf k}-{\bf k}^{\prime}}^2
\delta(\varepsilon_{{\bf k}^{\prime}}-\varepsilon_{{\bf
    k}}+\varepsilon^{h}_{{\bf k}^{\prime\prime}m}-\varepsilon^{h}_{{\bf
      k}^{\prime\prime}-{\bf k}+{\bf k}^{\prime}m^{\prime}}) 
\Big[ \hat{{\Lambda}}_{{\bf k},{\bf k}^{\prime}} \hat{\rho}^{>}_{{\bf
      k}^{\prime}} \hat{{\Lambda}}_{{\bf k}^{\prime},{\bf k}}
      \hat{\rho}^{<}_{{\bf k}} 
      |{\cal T}^{{\bf k}^{\prime\prime}m}_{{\bf k}^{\prime\prime}-{\bf
      k}+{\bf k}^{\prime}m^{\prime}}|^2 
      f^{h}_{{\bf k}^{\prime\prime}-{\bf k}+{\bf
      k}^{\prime}m^{\prime}} 
      \nonumber\\ && \mbox{\ \ } \times 
      \left(1 - f^{h}_{{\bf k}^{\prime\prime}m}\right) 
      - \hat{{\Lambda}}_{{\bf k},{\bf k}^{\prime}}
      \hat{\rho}^{<}_{{\bf 
      k}^{\prime}} \hat{{\Lambda}}_{{\bf k}^{\prime},{\bf k}}
      \hat{\rho}^{>}_{{\bf k}} 
      |{\cal T}^{{\bf k}^{\prime\prime}m}_{{\bf k}^{\prime\prime}-{\bf
      k}+{\bf k}^{\prime}m^{\prime}}|^2 
      \left(1 - f^{h}_{{\bf k}^{\prime\prime}-{\bf k}+{\bf k}^{\prime}m^{\prime}}\right) 
      f^{h}_{{\bf k}^{\prime\prime}m} 
 \Big]+ {\rm H.c.}, 
\label{scat_eh}
\end{eqnarray}
\begin{eqnarray}
\left. \partial_{t}\hat{\rho}_{{\bf k}}\right|_{\rm ex} &=& - \pi \sum_{{\bf
      k}^{\prime},{\bf
    k}^{\prime\prime},m,m^{\prime},\chi=\pm} 
\delta(\varepsilon_{{\bf k}^{\prime}}-\varepsilon_{{\bf
    k}}+\varepsilon^{h}_{{\bf k}^{\prime\prime}m}-\varepsilon^{h}_{{\bf
      k}^{\prime\prime}-{\bf k}+{\bf k}^{\prime}m^{\prime}}) 
\Big[ \hat{s}_{\chi} \hat{\rho}^{>}_{{\bf k}^{\prime}} \hat{s}_{-\chi} 
\hat{\rho}^{<}_{{\bf k}} 
|{\cal J}^{(\chi)\ {\bf k}^{\prime\prime}m}_{{\bf k}^{\prime\prime}-{\bf
      k}+{\bf k}^{\prime}m^{\prime}}|^2 
      f^{h}_{{\bf k}^{\prime\prime}-{\bf k}+{\bf
      k}^{\prime}m^{\prime}} 
      \left(1 - f^{h}_{{\bf k}^{\prime\prime}m}\right)
      \nonumber\\ &&\mbox{\ \ } - \hat{s}_{\chi} 
      \hat{\rho}^{<}_{{\bf k}^{\prime}} 
      \hat{s}_{-\chi} 
      \hat{\rho}^{>}_{{\bf k}} 
      |{\cal J}^{(\chi)\ {\bf k}^{\prime\prime}m}_{{\bf k}^{\prime\prime}-{\bf
      k}+{\bf k}^{\prime}m^{\prime}}|^2 
      \left(1 - f^{h}_{{\bf k}^{\prime\prime}-{\bf k}+{\bf k}^{\prime}m^{\prime}}\right)
      f^{h}_{{\bf k}^{\prime\prime}m} 
 \Big]+ {\rm H.c.}.
\label{scat_ex}
\end{eqnarray}
\end{widetext}
In these equations, $\hat{{\Lambda}}_{{\bf k},{\bf
  k}^{\prime}}=\hat{1} -i\lambda_c ({\bf k}\times{\bf
  k}^{\prime})\cdot\hat{\mbox{\boldmath$\sigma$\unboldmath}}$
describes the spin-mixing due to the conduction-valence band mixing which
originates from the EY spin relaxation mechanism.\cite{EY,opt-or}
Here $\lambda_c=\frac{\eta(1-\eta/2)}{3m_cE_g(1-\eta/3)}$ with 
$\eta=\frac{\Delta_{\rm SO}}{\Delta_{\rm SO} + E_g}$. $E_g$ and
  $\Delta_{\rm SO}$
are the band-gap and the spin-orbit
splitting of the valence band, respectively.\cite{opt-or} $n_i$ is the impurity
density. $Z_i$ is the charge number of the impurity which is taken
to be $Z_i=1$ throughout the paper.\cite{dislocation} $\varepsilon_{\bf k}=k^2/(2m_c)$
with $m_c$ being the conduction band effective mass. $V_{\bf q}$ is
the screened Coulomb potential where the screening is treated within
the random phase approximation (RPA),\cite{Mahan,wu-bap,hole} 
\begin{equation}
  V_{\bf q} = \frac{V^{(0)}_{\bf q}}{1-V^{(0)}_{\bf q}P^{(1)}({\bf
  q})}, 
\end{equation}
where
\begin{eqnarray}
  P^{(1)}({\bf q}) &=& \sum_{{\bf k},\sigma} \frac{f_{{\bf k}+{\bf
  q},\sigma}-f_{{\bf k},\sigma}}{\varepsilon_{{\bf k}+{\bf
  q}}-\varepsilon_{\bf k}} \nonumber\\ 
  &&\mbox{} + \sum_{{\bf k},m,m^{\prime}} |{\cal T}_{{\bf k},m}^{{\bf
  k}+{\bf q},m^{\prime}}|^2 \frac{f^{h}_{{\bf k}+{\bf
  q},m^{\prime}}-f^{h}_{{\bf k},m}}{\varepsilon^h_{{\bf k}+{\bf
  q},m^{\prime}}-\varepsilon^h_{{\bf k},m}}.\mbox{\ \ \ \ }
\end{eqnarray}
Here $V^{(0)}_{\bf q} = e^2/(\epsilon_0\kappa_0q^2)$ is the bare Coulomb
potential with $\epsilon_0$ and $\kappa_0$ representing the vacuum
permittivity and the static dielectric constant, respectively; 
$f_{{\bf k},\sigma}$ is the electron distribution on $\sigma$-spin
band; $f^{h}_{{\bf k},m}$ is the hole distribution on the hole $m$-spin
band. $\varepsilon^h_{{\bf
    k}m}=k^2/(2m^{\ast}_{m})$ stands for the hole energy dispersion with
$m^{\ast}_{m}$ being the hole effective mass. For heavy-hole
($m=\pm\frac{3}{2}$), $m^{\ast}_m=m_0/(\gamma_1-2\gamma_2)$; while for
light-hole ($m=\pm\frac{1}{2}$), $m^{\ast}_m=m_0/(\gamma_1+2\gamma_2)$.
Here $\gamma_1$ and $\gamma_2$ are the parameters of the
Luttinger-Kohn Hamiltonian in the spherical approximation\cite{hole} and $m_0$ is
the free electron mass. Note that
in these bands the spins are mixed.\cite{hole} Consequently, there are
form factors in the electron-hole Coulomb scattering: $|{\cal T}_{{\bf
    k}^{\prime} m^{\prime}}^{{\bf k} m}|^2 = |\langle \xi_{m}({\bf k})
| \xi_{m^{\prime}}({\bf k}^{\prime}) \rangle|^2$ where $ |
\xi_{m}({\bf k}) \rangle$'s are the eigen-states of the hole
Hamiltonian which can be found in
Ref.~\onlinecite{hole}. $M_{\lambda,{\bf q}}$ is the matrix element of
electron-phonon interaction with $\lambda$ being the phonon branch
index, which is further composed of the
electron--longitudinal-optical(LO)-phonon and
electron--acoustic-phonon interactions. The expressions of
$M_{\lambda,{\bf q}}$ can be found in
Ref.~\onlinecite{wu-3d}. $\omega_{\lambda {\bf q}}$ is the phonon 
energy spectrum. $N_{\lambda {\bf q}}^{\pm}=N_{\lambda {\bf
    q}}+\frac{1}{2}\pm\frac{1}{2}$ with $N_{\lambda {\bf q}}$ being
the phonon number at lattice temperature.

For electron-hole exchange scattering, the form factor $|{\cal
  J}^{(\chi)\ {\bf k}^{\prime\prime}m}_{{\bf k}^{\prime\prime}-{\bf 
    k}+{\bf k}^{\prime}m^{\prime}}|^2$ in Eq.~(\ref{scat_ex}) comes
from the electron-hole exchange interaction. The Hamiltonian for the
electron-hole exchange interaction consists of two parts: the
short-range part $H_{\rm SR}$ and the long-range part
$H_{\rm LR}$.\cite{opt-or,Sham} The short-range part contributes a term
proportional to $- \frac{1}{2} \frac{\Delta E_{\rm SR}}{|\phi_{\rm 3D}(0)|^2}
  \hat{\bf S}\cdot \hat{\bf s}$, 
where $\Delta E_{\rm SR}$ is the exchange splitting of the exciton ground
state and $|\phi_{\rm 3D}(0)|^2=1/(\pi a_0^3)$ with $a_0$ being the
exciton Bohr radius. $\hat{\bf S}$ and $\hat{\bf s}$ are the hole and
electron spin operators respectively. The long-range part gives a term
proportional to $\frac{3}{8} \frac{\Delta E_{\rm LT}}{|\phi_{\rm 3D}(0)|^2}
  (\hat{M}_1 \hat{1} + \hat{M}_z \hat{s}_z + \frac{1}{2}\hat{M}_{-} \hat{s}_{+} +
  \frac{1}{2}\hat{M}_{+} \hat{s}_{-} )$ 
where $\Delta E_{\rm LT}$ is the longitudinal-transverse splitting; $\hat{M}_1$,
$\hat{M}_z$, $\hat{M}_{-}$, and $\hat{M}_{+}(=\hat{M}_{-}^{\dagger})$
are operators in hole spin space. $\hat{s}_{\pm} =
\hat{s}_x \pm i \hat{s}_y$ are the electron spin ladder operators. The
expressions for $\hat{M}$'s can be obtained from the
expressions of $H_{\rm LR}$ in Ref.~\onlinecite{Sham}. Specifically, 
the spin-flip matrix $\hat{M}_{-}$ in the spin unmixed base can be
written as (in the order of $|\frac{3}{2}\rangle$,
$|\frac{1}{2}\rangle$, $|-\frac{1}{2}\rangle$, $|-\frac{3}{2}\rangle$), 
\begin{eqnarray}
  \hat{M}_{-} = \frac{1}{K^2}
  \begin{bmatrix} 
    0 & 0 & 0 & 0 \\[8pt] 
    - \frac{2}{\sqrt{3}} K_{\parallel}^2 & \frac{4}{3} K_z K_{-} &
      \frac{2}{3} K_{-}^2 & 0 \\[8pt]
      \frac{4}{\sqrt{3}}K_z K_{+} & -\frac{8}{3} K_z^2 & -\frac{4}{3}
      K_z K_{-} & 0 \\[8pt]
      2 K_{+}^2 & -\frac{4}{\sqrt{3}} K_z K_{+} & -\frac{2}{\sqrt{3}}
      K_{\parallel}^2 & 0 
  \end{bmatrix}.
\end{eqnarray}
Here ${\bf K}={\bf k}^{\prime}+{\bf k}^{\prime\prime}$ is the
center-of-mass momentum of the interacting electron-hole
pair. $K_{\pm} = K_x \pm i K_y$ and $K_{\parallel}^2 = K_x^2 +
K_y^2$. $\hat{M}_1 \hat{1} + \hat{M}_z \hat{s}_z$ corresponds
to the spin-conserving process which is irrelevant 
in the study of spin relaxation. Summing up the
contribution from the short-range and long-range parts and
keeping only the spin-flip terms, one obtains 
\begin{equation}
{\cal J}_{{\bf k}^{\prime}-{\bf q} m^{\prime}}^{(\pm)\ {\bf k}^{\prime} m} =
\langle \xi_{m}({\bf k}^{\prime}) | {\cal J}^{(\pm)} |
\xi_{m^{\prime}}({\bf k}^{\prime}-{\bf q}) \rangle, 
\end{equation}
with ${\cal J}^{(\pm)} = \left[ \frac{3}{16} \frac{\Delta
    E_{\rm LT}}{|\phi_{\rm 3D}(0)|^2} \hat{M}_{\pm} - \frac{1}{4} \frac{\Delta
    E_{\rm SR}}{|\phi_{\rm 3D}(0)|^2} \hat{S}_{\pm}
    \right]$. $\hat{S}_{\pm}=\hat{S}_x\pm i\hat{S}_y$ are
the hole spin ladder operators. It should be noted that in GaAs, 
$\Delta E_{\rm LT}=0.08$~meV is four times as large as $\Delta
E_{\rm SR}=0.02$~meV.\cite{para,bappara1} Thus in GaAs the long-range
part of electron-hole exchange
interaction should be more important than the short-range
part. Later in this paper, we will compare the contributions from the 
short-range and long-range parts to show that the long-range part
is much more important than the short-range part in GaAs. 
We will also compare the results from the fully microscopic KSBE approach
with those from the analytical formula widely used in the
literature.\cite{opt-or,Song,Fishman,Zerrouati,spintronics}

In summary, all relevant spin relaxation mechanisms have been fully
incorporated in our KSBE approach. By solving the KSBEs
[Eq.~(\ref{KSBE})], one obtains the temporal evolution of spin density
matrix $\hat{\rho}_{\bf k}$. After that, the time evolution of 
macroscopic quantities, such as, the electron spin density along
the $z$-axis, $s_z(t)=\sum_{\bf k}{\rm Tr}\left[\hat{s}_z\hat{\rho}_{\bf
    k}(t)\right]$, are obtained. By fitting the decay of $s_z$ with an
exponential decay, one obtains the SRT. The SRTs under various
conditions are studied with the underlying physics 
discussed. The material parameters used are listed in Table I. The 
parameter for strain-induced SOC $\beta$ is always taken to be zero
unless otherwise specified. The numerical scheme is laid out in
Appendix~A. The error of our computation is less than $5~\%$ according
to our test.

We first compare our calculation with experiments. In Appendix~B, we
compare the SRTs calculated from the KSBEs with those from experiments in 
Refs.~\onlinecite{Awsch98}, \onlinecite{Nex-decr}
and \onlinecite{Zerrouati}. We find good agreement with experimental
data in almost the whole temperature or photo-excitation density 
range with only one fitting
parameter $\gamma_{\rm D}$.\cite{listsoc} This demonstrates that our
calculation has achieved {\em quantitative accuracy} in modeling the
spin relaxation in metallic regime.

\begin{table}[htbp]
\caption{Material parameters used in the calculation (from
  Ref.~\onlinecite{para} unless otherwise specified)}
\begin{tabular}{llllllllll}\hline\hline
  \mbox{}      &\mbox{}\mbox{}\mbox{}&  GaAs  &\mbox{}\mbox{}&  GaSb
  &\mbox{}\mbox{}&  InAs  &\mbox{}\mbox{}&   InSb  \\
\hline
 $E_g$ (eV)      &\mbox{}\mbox{}\mbox{}&  1.52  &\mbox{}\mbox{}&  0.8113
  &\mbox{}\mbox{}&   0.414  &\mbox{}\mbox{}&   0.2355  \\   
$\Delta_{\rm SO}$ (eV) &\mbox{}\mbox{}\mbox{}&  0.341  &\mbox{}\mbox{}&  0.75
  &\mbox{}\mbox{}&   0.38  &\mbox{}\mbox{}&   0.85      \\
$m_c/m_0$          &\mbox{}\mbox{}\mbox{}&  0.067  &\mbox{}\mbox{}&  0.0412
  &\mbox{}\mbox{}&   0.023  &\mbox{}\mbox{}&   0.0136      \\
$\kappa_0$         &\mbox{}\mbox{}\mbox{}&  12.9  &\mbox{}\mbox{}&  15.69
  &\mbox{}\mbox{}&   15.15  &\mbox{}\mbox{}&   16.8      \\
$\kappa_{\infty}$  &\mbox{}\mbox{}\mbox{}&  10.8  &\mbox{}\mbox{}&  14.44
  &\mbox{}\mbox{}&  12.25  &\mbox{}\mbox{}&   15.68      \\
$\omega_{\rm LO}$ (meV) &\mbox{}\mbox{}\mbox{}&  35.4  &\mbox{}\mbox{}& 28.95
  &\mbox{}\mbox{}&   27.0 &\mbox{}\mbox{}&   23.2      \\
$v_{sl}$ ($10^3$~m/s)    &\mbox{}\mbox{}\mbox{}&  5.29  &\mbox{}\mbox{}&  4.01
  &\mbox{}\mbox{}&    4.28   &\mbox{}\mbox{}&   3.4081      \\    
$v_{st}$ ($10^3$~m/s)    &\mbox{}\mbox{}\mbox{}&  2.48  &\mbox{}\mbox{}&  2.4
  &\mbox{}\mbox{}&   1.83  &\mbox{}\mbox{}&   2.284      \\   
$D$  ($10^3$~kg/m$^{3}$)   &\mbox{}\mbox{}\mbox{}& 5.31 &\mbox{}\mbox{}&  5.6137
  &\mbox{}\mbox{}&    5.9   &\mbox{}\mbox{}&   5.7747      \\    
$\varXi$ (eV)    &\mbox{}\mbox{}\mbox{}&  8.5  &\mbox{}\mbox{}&  8.3
  &\mbox{}\mbox{}&  5.8   &\mbox{}\mbox{}&   14.0     \\
$e_{14}$ (V/m)    &\mbox{}\mbox{}\mbox{}&  1.41$\times 10^{9}$ &\mbox{}\mbox{}&  9.5$\times 10^8$  
&\mbox{}\mbox{}&  0.35$\times 10^9$   &\mbox{}\mbox{}&   4.7$\times 10^8$  \\
$\gamma_{\rm D}$ (eV$\cdot$\AA$^3$) &\mbox{}\mbox{}\mbox{}&  23.9$^{a,b}$ &\mbox{}\mbox{}&  168$^{b}$ &\mbox{}\mbox{}&   42.3$^{b}$   &\mbox{}\mbox{}&   389$^{b}$  \\
$\Delta E_{\rm LT}$ (meV)   &\mbox{}\mbox{}\mbox{}&  0.08$^{c}$  &\mbox{}\mbox{}&  0.02$^{d}$
  &\mbox{}\mbox{}&   ---    &\mbox{}\mbox{}&   ---  \\
$\Delta E_{\rm SR}$ (meV)   &\mbox{}\mbox{}\mbox{}&  0.02$^{c}$  &\mbox{}\mbox{}&  0.024$^{e}$
  &\mbox{}\mbox{}&   ---    &\mbox{}\mbox{}&   ---  \\
$\gamma_1$         &\mbox{}\mbox{}\mbox{}&  6.85  &\mbox{}\mbox{}&  11.8
  &\mbox{}\mbox{}&   19.67    &\mbox{}\mbox{}&   35.08  \\
$\gamma_2$         &\mbox{}\mbox{}\mbox{}&  2.5$^f$  &\mbox{}\mbox{}&  4.65$^f$
  &\mbox{}\mbox{}&   8.83$^f$  &\mbox{}\mbox{}&   16.27$^f$  \\
\hline\hline
\end{tabular}\\
 $^a$ Ref.~\onlinecite{wu-soc}. \quad\quad $^b$ Ref.~\onlinecite{soc35}.  \quad\quad 
$^c$ Ref.~\onlinecite{bappara1}. \quad\quad $^d$
  Ref.~\onlinecite{bappara2}. \quad\quad $^e$
  Ref.~\onlinecite{Aronov}. \quad\quad $^f$ Obtained from
  Ref.~\onlinecite{para} by $\frac{1}{2}(\gamma_2+\gamma_3)$
\end{table}

In the rest part of this section, we briefly comment on the merits
of the fully microscopic KSBE approach. First, for the DP mechanism. 
Previously, the widely used analytical formula for the SRT due to the DP
mechanism is\cite{opt-or,spintronics} 
\begin{equation}
  \frac{1}{\tau_{\rm DP}} = Q \alpha^2 \frac{\left(k_{\rm B}
  T\right)^3}{E_g} \tau_p, 
\label{DPapp}
\end{equation}
where $Q$ is a numerical constant which varies around 1 depending on
the dominant momentum scattering mechanism, $\tau_p$ is
the momentum relaxation time, and $\alpha=2\gamma_{\rm D}\sqrt{2 m_c^{3} E_g}$
is a dimensionless constant. This formula is derived from 
\begin{equation}
\label{tau-taup}
\tau_{\rm DP}^{-1}=\langle (|{\bf \Omega}({\bf k})|^2-\Omega_z^2({\bf k}))
\tau_p({\bf k}) \rangle
\end{equation}
[$\tau_p({\bf k})$ is the momentum relaxation time of the state with momentum
  ${\bf k}$]  by performing ensemble averaging over the Boltzmann
distribution.\cite{spintronics} An important fact about this formula is
that it is derived in the elastic scattering approximation, which
artificially confines the random-walk-like spin precession due to the
${\bf k}$-dependent spin-orbit field only within
the same energy states. However, in the genuine case the inelastic
electron-phonon scattering (especially the electron--LO-phonon
scattering) as well as the carrier-carrier Coulomb scattering can be
more important, and the random spin precession (the ``inhomogeneous
broadening'') should be fully counted for the whole ${\bf k}$-space,
instead only within the same energy states. On the other hand, the KSBE
approach, which solves the spin precession and the momentum
scattering self-consistently, takes full account of the inhomogeneous
broadening and the counter effects of scattering on the inhomogeneous broadening.
Moreover, the electron-electron scattering
(although it does not contribute to the mobility) as well as the electron-hole
scattering, which have been demonstrated to have important effects on
spin relaxation/dephasing in two-dimensional systems,\cite{wu-early,wu-review,Harley,Ivchenko,Ji,lowT,highP,hot-e,multi-band,multi-valley,wu-hole,wu-bap,terahertz,Lai} are always neglected in previous
studies based on Eq.~(\ref{DPapp}). On the contrary, the fully
microscopic KSBE approach includes both the electron-electron
and electron-hole Coulomb scatterings. In this work, we will discuss the
effect of the electron-electron and electron-hole Coulomb scatterings on spin relaxation.

Second, for the EY mechanism. The formula commonly used in the literature
for the SRT due to the EY mechanism reads\cite{opt-or,spintronics} 
\begin{equation}
  \frac{1}{\tau_{\rm EY}} = A \left(\frac{k_{\rm B}T}{E_g}\right)^2 \eta^2
  \left(\frac{1-\eta/2}{1-\eta/3}\right)^2 \frac{1}{\tau_p},
\label{EYapp}
\end{equation}
where the numerical factor $A$ is of order 1 depending on the dominant
scattering mechanism. This formula, which is also based on the elastic
scattering approximation, has similar problems with those of
Eq.~(\ref{DPapp}). On the contrary, the fully microscopic KSBE approach,
which incorporates {\em all} the EY spin-flip processes in {\em all}
relevant scatterings,\cite{EY_pre} fully takes into account of the
spin relaxation due to the EY mechanism.

Third, for the BAP mechanism. The commonly used formula for the SRT due to the BAP
mechanism for non-degenerate holes (assuming all holes are free) is\cite{spintronics}
\begin{equation}
  \frac{1}{\tau_{\rm BAP}} = \frac{2}{\tau_0} n_h a_{\rm B}^3 \frac{\langle v_{\bf
  k}\rangle}{v_{\rm B}}, 
\label{BAPapp1}
\end{equation}
where $a_{\rm B}$ is the exciton Bohr radius, $1/\tau_0=(3\pi/64)\Delta
E_{\rm SR}^2 / E_{\rm B}$ with $E_{\rm B}$ being the exciton Bohr energy, $n_h$ is the
hole density, $\langle v_{\bf k}\rangle = \langle k/m_c\rangle$
is the average electron velocity, and 
$v_{\rm B}=1/(m_R a_{\rm B})$ with $m_R\approx m_c$ being the reduced mass of
the electron-hole pair. For degenerate holes,\cite{spintronics}
\begin{equation}
  \frac{1}{\tau_{\rm BAP}} = \frac{3}{\tau_0} n_h a_{\rm B}^3 \frac{\langle v_{\bf
  k}\rangle}{v_{\rm B}} \frac{k_{\rm B} T}{E_{\rm F}^{h}}, 
\label{BAPapp2}
\end{equation}
with $E_{\rm F}^{h}$ denoting the Fermi energy of holes. All these formulae are
obtained within the elastic scattering approximation. Previously, it
has been shown that for two-dimensional system the elastic scattering
approximation is invalid for low temperature regime where the Pauli
blocking of electrons is important.\cite{wu-bap}
In bulk system, similar problem also exists.\cite{Amo} However, the KSBE
approach keeps all the electron-hole exchange scattering terms, and
thus gives correct results.

Finally, we point out that for excitation far away from the equilibrium,
such as excitation with high spin polarization, the spin-conserving
scatterings are very important for redistribution of electrons in each
spin band, which also affects the spin dynamics. This effect is
automatically kept in our approach, but missing in the analytical formulae.

Although the analytical formulae based on single particle
approach [Eqs.~(\ref{DPapp})-(\ref{BAPapp2})]
have some disadvantages, for non-degenerate electron system they still
give qualitatively good results. However, quantitative analysis based
on them should be questioned. It has already been extensively
demonstrated in two-dimensional system (both
theoretically\cite{wu-early,highP,hot-e,multi-band,wu-hole,lowT,wu-bap,multi-valley,terahertz,Ivchenko,wu-review,nonMarkov,wu-strain}
and
experimentally\cite{Harley,Ji,Lai,wu-exp-hP,tobias,wu-exp,Awsch-exp,Zheng,Zheng2,Cui,Harleyexp1,Harleyexp2,Harleyexp3,Amo})
that the above single-particle approach, with even the recent
developments which exert closer examination on both the electron
distribution and the energy-dependent scattering cross
sections, are inadequate in accounting for the spin
relaxation.\cite{harleybook} The same is true for bulk
system.\cite{GaAs-cal1,GaAs-cal2,GaAs-cal3,LO,Maialle}
Nevertheless, when the electron-impurity scattering is much stronger than the
electron-electron (electron-hole) and electron-phonon scatterings
[i.e., in degenerate electron (hole) regime in $n$-type ($p$-type) bulk
  semiconductors with low excitation density], the above approach for
the DP and EY spin relaxations and the later
developments\cite{GaAs-cal1,GaAs-cal2,GaAs-cal3} are applicable, thanks
to the fact that the impurity density is larger than or equal to the
electron (hole) density in $n$-type ($p$-type) bulk semiconductors
(which is quite different from the two-dimensional system). On
the other hand, the fully microscopic KSBE approach takes full
account of the DP, EY and BAP spin relaxations by solving the kinetic
equation directly. It is applicable in all parameter regimes in
the metallic regime. Furthermore, the fully microscopic KSBE approach
can be applied to system {\em far away} from equilibrium, such as under
strong electric field where the hot-electron effect is
prominent,\cite{hot-e,multi-valley,terahertz} or with high spin
polarization, and/or hot photo-carriers.\cite{highP,wu-exp-hP} It is even valid
  for system out of the motional narrowing regime, or in the non-Markovian
limit.\cite{wu-hole,nonMarkov}

\section{spin relaxation in $n$-type III-V semiconductors}

\subsection{Comparison of different spin relaxation mechanisms}

In $n$-type III-V semiconductors, the BAP mechanism is ineffective due
to the lack of holes. The remaining mechanisms are the DP and EY
mechanisms. Previously, it is widely accepted in the literature that
the EY mechanism dominates spin relaxation at low temperature, while
the DP mechanism is important at high
temperature.\cite{opt-or,Song,spintronics} Most studies concerned are based
on the analytical formulae [Eqs.~(\ref{DPapp}) and (\ref{EYapp})],
which give
\begin{equation}
  \frac{\tau_{\rm EY}}{\tau_{\rm DP}} =  k_{\rm B} T
  \tau_p^2 \frac{Q\alpha^2 E_g}{A\eta^2}
  \left(\frac{1-\eta/3}{1-\eta/2}\right)^2.  
\label{EYDP1}
\end{equation}
From the above equation, one arrives at the conclusion that the EY
mechanism is dominant at low temperature and/or when the momentum
scattering is strong (such as, for heavily doped samples). However,
the electron system may enter into the degenerate regime at low
temperatures and/or in heavily doped samples, in which the above
conclusion fails. A revised expression is obtained by replacing $k_{\rm B}T$
with the average kinetic energy $\langle \varepsilon_{\bf k}\rangle$,\cite{spintronics} 
\begin{equation}
  \frac{\tau_{\rm EY}}{\tau_{\rm DP}} = \frac{2Q}{3A}\langle \varepsilon_{\bf k}\rangle
  \tau_p^2 \Theta,
\label{EYDP2}
\end{equation}
with $\Theta = \frac{\alpha^2 E_g}{\eta^2}
\left(\frac{1-\eta/3}{1-\eta/2}\right)^2$. This is still correct only
qualitatively.

\begin{table}[htbp]
\caption{The factor $\Theta$ for III-V semiconductors}
\begin{tabular}{lllllllllllll}\hline\hline
 \mbox{}      &\mbox{}\mbox{}\mbox{}&  GaAs  &\mbox{}\mbox{}&  GaSb
  &\mbox{}\mbox{}&  InAs  &\mbox{}\mbox{}&   InSb &\mbox{}\mbox{}&   InP \\
\hline
$\Theta$ (eV) &\mbox{}\mbox{}\mbox{}&  0.23  &\mbox{}\mbox{}&  0.12
  &\mbox{}\mbox{}&  3.6$\times 10^{-4}$  &\mbox{}\mbox{}&   9.2$\times 10^{-4}$  &\mbox{}\mbox{}&  0.27$^a$\\  
\hline\hline
\end{tabular}\\
$^a$ The SOC parameter $\gamma_{\rm D}$ is from Ref.~\onlinecite{soc35}, \\
other parameters are from Ref.~\onlinecite{para}. 
\end{table}

\begin{figure}[htb]
\includegraphics[height=5.cm]{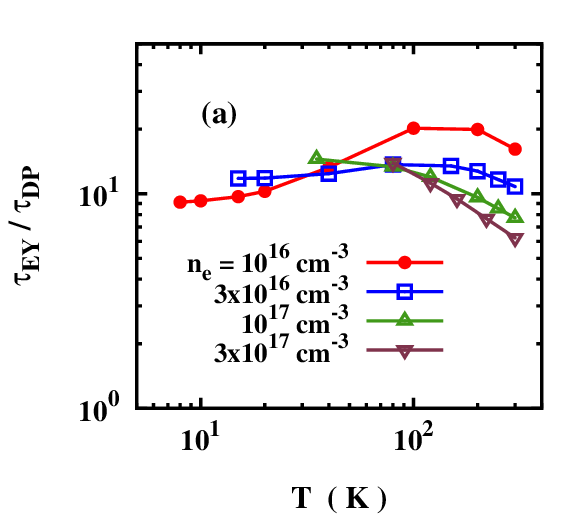}
\includegraphics[height=5.cm]{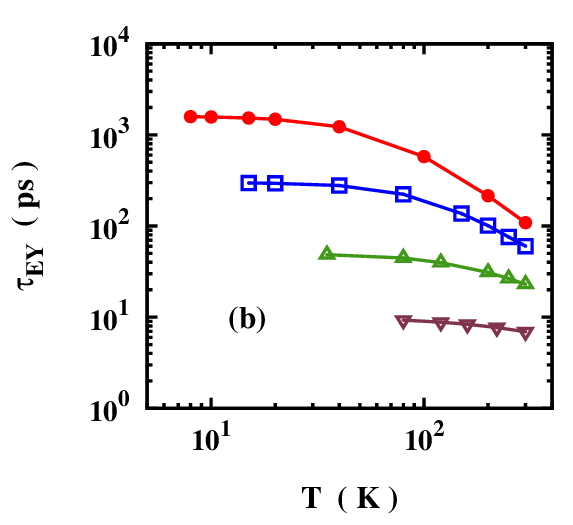}
\includegraphics[height=5.cm]{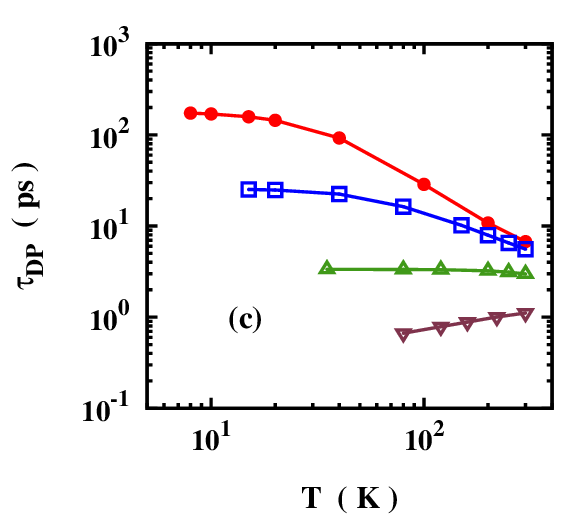}
\caption{(Color online) $n$-InSb. Ratio of the SRT due to the EY mechanism
  $\tau_{\rm EY}$ to that due to the DP mechanism $\tau_{\rm DP}$ (a),
  $\tau_{\rm EY}$ (b) and $\tau_{\rm DP}$ (c) as
  function of temperature for various electron densities.
  $n_{e}=10^{16}$~cm$^{-3}$ (curve with $\bullet$), $3\times
  10^{16}$~cm$^{-3}$ (curve with $\square$), $10^{17}$~cm$^{-3}$
  (curve with $\triangle$), $3\times 10^{17}$~cm$^{-3}$
  (curve with $\triangledown$). The corresponding Fermi temperatures for
  those densities are $T_{\rm F}=144$, 300, 670, and 1390~K respectively. $n_i=n_{e}$.}
\label{fig:EYINSB}
\end{figure}

In this work, we reexamine the problem from the fully microscopic KSBE
approach. Let us first examine the case of InSb. InSb is a narrow
band-gap semiconductor where the spin relaxation is believed to be
dominated by the EY mechanism at low temperature
previously.\cite{Chazalviel,Song,spintronics}
The factor $\Theta$ is much smaller than that of GaAs, which indicates
the importance of the EY mechanism according to Eq.~(\ref{EYDP2}) (A
list of the factor $\Theta$ for different materials is given in Table II). In
Fig.~\ref{fig:EYINSB}(a), we plot the ratio of the SRT due to the EY
mechanism, $\tau_{\rm EY}$, to that due to the DP mechanism, $\tau_{\rm DP}$,
calculated from the KSBEs, as function of temperature for various electron
densities. In the calculation, $n_i\equiv n_{e}$. Remarkably, it is noted
that the ratio $\tau_{\rm EY}/\tau_{\rm DP}$ is always larger than 1, and in
most cases it is even larger than 10, i.e., the spin relaxation in
$n$-type InSb is {\em not} dominated by the EY mechanism.
Moreover, the temperature dependence is not monotonic, which is
different from the intuition given by Eq.~(\ref{EYDP2}). This
can be understood as following: The increase of the ratio comes from
the increase of $\langle \varepsilon_{\bf k}\rangle$ which is
understood easily. The decrease of the ratio comes from the decrease
of $\tau_p$ due to the increase of the electron--LO-phonon scattering
with temperature which becomes important for $T\gtrsim
80$~K in InSb. Therefore, after a crossover regime, the ratio eventually
decreases with temperature. Moreover, for large electron
density $n_{e}\ge 10^{17}$~cm$^{-3}$, the electron system is in the
degenerate regime and $\langle \varepsilon_{\bf k}\rangle$ changes
slowly with temperature which further facilitates the decrease of the
ratio. To elucidate it clearly, we plot the SRT due to the EY mechanism
and DP mechanism in Fig.~\ref{fig:EYINSB}(b) and (c) respectively. It
is seen from Fig.~\ref{fig:EYINSB}(b) that the SRT due
to the EY mechanism always decreases with temperature due to both the increase of
$\langle \varepsilon_{\bf k}\rangle$ and the enhancement of the 
scattering. From Fig.~\ref{fig:EYINSB}(c), it is further noted that
the temperature dependence of the SRT due to the DP mechanism
is, however, different for low density and high
density regimes: in low density regime, the SRT decreases with the
temperature, whereas in high density regime it increases.
The decreases of SRT with temperature at low density is consistent
with previous studies in the literature.\cite{DP,Song}
The increases of SRT with temperature at high density, however,
is because that the electron--LO-phonon scattering increases with
temperature faster than the inhomogeneous broadening $\sim \langle
(|{\bf \Omega}({\bf k})|^2-\Omega_z^2({\bf k}))\rangle \propto \langle
\varepsilon_{\bf k}^3\rangle$ in the degenerate regime. Note
 that the increase of the SRT with temperature at such high temperature
($T>100$~K) without any magnetic field has neither been observed in
  experiment nor been predicted in theory.
Note that the results for high
density cases are only given at high temperatures due to the limitation
of our computation power. However, according to the above analysis,
the ratio in the low temperature regime
should be larger than its value at 300~K for high density cases.
For low density cases with temperature lower than our calculation
range, both the momentum scattering and the inhomogeneous broadening
change little with temperature as the electron system is
degenerate. Therefore the ratio $\tau_{\rm EY}/\tau_{\rm DP}$
changes slowly. Consequently, it does not
change the conclusion that the EY mechanism is less efficient than the
DP mechanism in $n$-type InSb.\cite{note1}  In contrast, based on 
Eq.~(\ref{EYDP2}),
Song and Kim reported that the EY mechanism is more important than the
DP mechanism for temperature lower than 5~K.\cite{Song}

What makes our conclusion different from that in the literature is
that most of the previous investigations use Eqs.~(\ref{DPapp}) and
(\ref{EYapp}) to calculate $\tau_{\rm DP}$ and $\tau_{\rm EY}$.\cite{Song,spintronics}
However, these equations are only applicable for non-degenerate
electron system. If it is used in the degenerate electron system,
it {\em exaggerates} the relative efficiency of the EY mechanism according
to Eq.~(\ref{EYDP1}).

\begin{figure}[htb]
\includegraphics[height=5.cm]{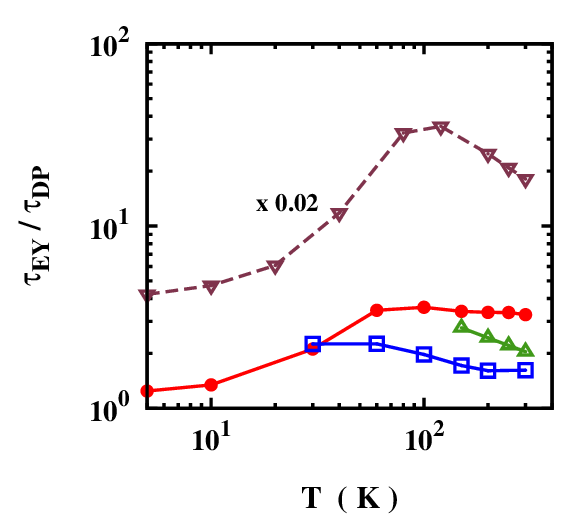}
\caption{(Color online) $n$-InAs (solid curves) and $n$-GaAs (dashed curve).
  Ratio of the SRT due to the EY mechanism $\tau_{\rm EY}$ to that due to
  the DP mechanism $\tau_{\rm DP}$ as function of temperature for various
  electron densities: $n_{e}=10^{16}$~cm$^{-3}$ (curve
  with $\bullet$), $2\times 10^{17}$~cm$^{-3}$ (curve with
  $\square$), $10^{18}$~cm$^{-3}$ (curve with $\triangle$) for
  InAs, and $n_{e}=10^{16}$~cm$^{-3}$ (curve with $\triangledown$) for
  GaAs (note that the value in the figure has been rescaled by a
  factor of 0.02). The Fermi temperature $T_{\rm F}$ is 85, 629, and 1840~K
  for InAs as well as 29~K for GaAs respectively. $n_i=n_{e}$.} 
\label{fig:EYINAS}
\end{figure}

We further examine other two III-V semiconductors: InAs and GaAs. 
Figure~\ref{fig:EYINAS} shows the ratio $\tau_{\rm EY}/\tau_{\rm
  DP}$ for different electron densities as function of temperature for
both InAs and GaAs. It is seen that the ratio is always larger than 1
for both InAs\cite{note2} and GaAs,\cite{note9} which is different from previous
results in the literature.\cite{Litvinenko,Song} Especially, for GaAs
the ratio is larger than 100, which indicates that the EY mechanism is
{\em irrelevant} for $n$-type GaAs. The temperature dependence of the
ratio is similar to that in InSb and the underlying physics is also
the same.

In summary, we find that {\em the EY mechanism is less efficient than the DP
  mechanism in $n$-type GaAs, InAs, and InSb}. According to the data
listed in Table II, we believe that the same conclusion holds for
other $n$-type III-V semiconductors.

\subsection{DP spin relaxation}
As the BAP mechanism and the EY mechanism are unimportant for spin
relaxation in $n$-type III-V semiconductors, in this subsection we
focus on spin relaxation due to the DP mechanism. The effect of
the electron-electron Coulomb scattering on spin relaxation is
studied. Previously, it was found that the electron-electron  
scattering plays an important role in two-dimensional
system.\cite{wu-early,wu-hole,hot-e,highP,lowT,multi-band,multi-valley,wu-bap,terahertz,Franz}
Especially, it becomes the dominant scattering mechanism in high
mobility samples at low temperature, where other scattering mechanisms
are relatively weak.\cite{wu-exp-hP,Ivchenko,lowT,Franz} 
Consequently  $\tau_p({\bf k})$ in Eq.\ (\ref{tau-taup})
 should be replaced by $\tau_p^{\ast}({\bf k})$ with
$1/\tau_p^{\ast}({\bf k})$ including the electron-electron
scattering $1/\tau_p^{ee}({\bf k})$ [later we will use the symbols
  $\tau_p^{\ast}$ and $\tau_p^{ee}$ to denote the ensemble averaged 
value].\cite{wu-early,highP,Ivchenko,lowT}
Remarkably, the SRT due to
the electron-electron scattering has {\em nonmonotonic} temperature
dependence: in the low temperature (degenerate) regime the SRT
increases with temperature as the electron-electron scattering does;
in the high temperature (non-degenerate) regime the electron-electron
scattering decreases with the temperature, so does the
SRT.\cite{Ivchenko,lowT} Thus there is a peak $T_c$ in the temperature
dependence of the SRT which is comparable with the Fermi
temperature.\cite{lowT,Franz} This prediction was confirmed by
experiments very recently.\cite{Ji} However, in bulk
semiconductors, the role of electron-electron scattering in spin
relaxation is still unclear. Although Glazov and Ivchenko have
discussed the problem, they only gave an approximate expression of the
SRT due to the electron-electron scattering in the non-degenerate regime,
while the relative importance of the electron-electron scattering
compared with other scattering
mechanisms is not addressed.\cite{Ivchenko} In this subsection, we
present a close study on the effect of the electron-electron scattering on
spin relaxation in bulk semiconductors. We use GaAs as an example,
where the behavior is applicable to all III-V semiconductors given that
the system is in the motional narrowing regime.

\begin{figure}[htb]
\includegraphics[height=5.cm]{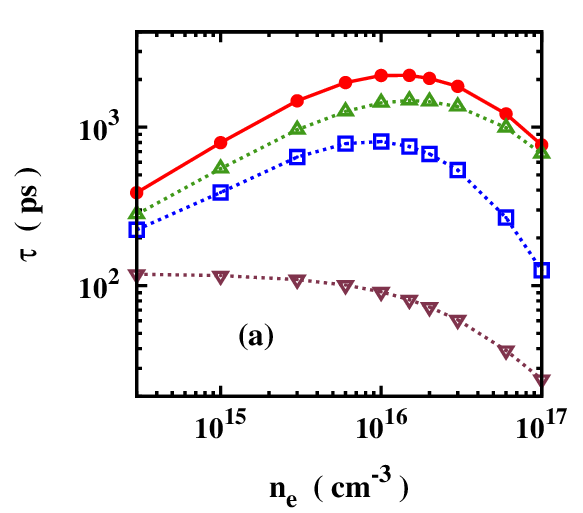}
\includegraphics[height=5.cm]{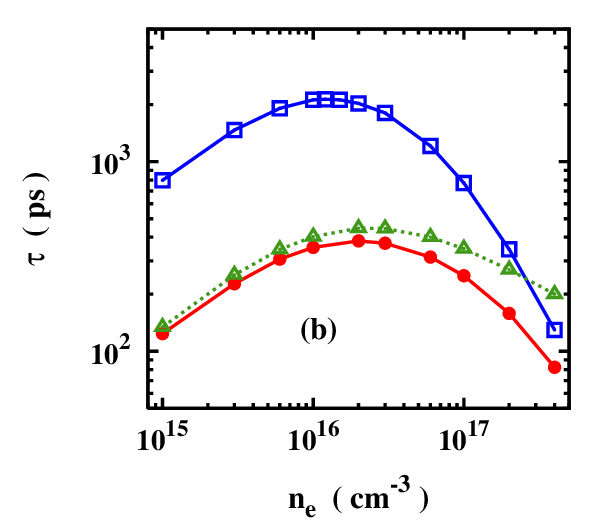}
\caption{(Color online) $n$-GaAs at $T=40$~K. (a) SRT $\tau$ as
  function of electron density $n_{e}$ ($n_i=n_{e}$) from full calculation
  (curve with $\bullet$), from calculation with only the electron-electron
  scattering (curve with $\square$), with only the
  electron-impurity scattering (curve with $\triangle$), and with
  only the electron-phonon scattering (curve with $\triangledown$);
  (b) SRT $\tau$ as
  function of electron density $n_{e}$ ($n_i=n_{e}$) for the case with
  strain (curve with $\bullet$: with both the linear and the cubic SOC; curve
  with $\triangle$: with only the linear SOC) and the case without strain
  (curve with $\square$).}
\label{fig:ne}
\end{figure}

\subsubsection{Electron density dependence}

We first discuss the electron density dependence of the SRT. It is
noted that during the variation of electron density, the impurity
density also varies as $n_i=n_{e}$. This is different from the situation in
2DES, where the impurity density can be different from the electron
density due to the modulation doping. In Fig.~\ref{fig:ne}(a), we plot the SRT as function of
electron density for $T=40$~K. It is noted that, remarkably, the
density dependence of the SRT is {\em nonmonotonic} and there is a
peak in the $\tau$-$n_{e}$ curve. Previously, the nonmonotonic density
dependence of the SRT was observed in low temperature ($T\le 5$~K)
measurements, where the
localized electrons play an important role and the electron system is
in the Mott metal-insulator transition area.\cite{opt-nmr,localize,localize2,Awsch98,localizeReview}
The localized electrons have different spin relaxation
mechanisms 
and the scatterings of the localized electrons and free electrons give
rise to the nonmonotonic density dependence.\cite{localize,localizeReview}
It is noted that, up till now, there is no report on the nonmonotonic density
dependence in the metallic regime in $n$-type bulk III-V
semiconductors. It should be further pointed out that the nonmonotonic
density dependence and the appearance of the peak is a {\em universal}
behavior in the metallic regime in $n$-type bulk III-V
semiconductors.

We also plot the SRTs calculated with only the
electron-electron scattering (curve with $\square$), with only
the electron-impurity scattering (curve with $\triangle$), and with
only the electron-phonon scattering (curve with $\triangledown$) in
the figure to elucidate the role of these scatterings in
spin dynamics. It is noted that the electron-impurity and
electron-electron scatterings are the relevant scattering
mechanisms, while the electron-phonon scattering is much weaker than
the two as the temperature is low. Interestingly, both the
electron-electron scattering and the electron-impurity scattering lead to
nonmonotonic density dependence of SRT.

For the SRT due to the electron-electron scattering, the nonmonotonic
behavior comes from the nonmonotonic density dependence of the
electron-electron scattering. The density and temperature dependences
of the electron-electron scattering have been investigated in
spin-unrelated problems.\cite{Vignale} From these works,
after some approximation, the asymptotic density and temperature
dependences of the electron-electron scattering time
$\tau_p^{ee}$ in the degenerate and non-degenerate regimes
are given by,\cite{Vignale,Ivchenko}
\begin{eqnarray}
  \tau_p^{ee} \propto n_{e}^{2/3}/ T^2 \quad \quad {\rm for}\quad  T\ll T_{\rm F}, 
\label{taupee1} \\
  \tau_p^{ee} \propto T^{3/2} / n_{e} \quad\quad {\rm for}\quad  T\gg T_{\rm F}.
\label{taupee2}
\end{eqnarray}
From above equations, one notices that the electron-electron scattering
in the non-degenerate and degenerate regimes has different density
and temperature dependence. In the non-degenerate (low
density) regime the electron-electron scattering increases with
electron density, while the inhomogeneous broadening
$\sim \langle (|{\bf \Omega}({\bf  k})|^2-\Omega_z^2({\bf k}))\rangle$
changes slowly as the distribution function is close to the Boltzmann
distribution. The SRT thus increases with
density. In degenerate (high density) regime, both $\tau_p^{ee}$ and
the inhomogeneous broadening increases with electron density. Thus,
the SRT, which can be estimated as $\tau \sim 1/[\langle |{\bf
  \Omega}({\bf k})|^2-\Omega_z^2({\bf k}) \rangle
  \tau_p^{ee}]$,\cite{Ivchenko,lowT} decreases with density.

For the SRT due to the electron-impurity scattering, the scenario is
similar: In the non-degenerate regime, the distribution function
is close to the Boltzmann distribution. The inhomogeneous broadening
hence changes slowly with density. The electron-impurity scattering
increases as $\propto n_i \langle V_{q}^2 \rangle$, which increases
with the impurity density because $\langle V_{q}^2 \rangle$ changes
little with the density when the distribution is close to the
Boltzmann distribution. The SRT hence increases with density. In the
degenerate regime, the inhomogeneous broadening increases as
$\sim\langle (|{\bf \Omega}({\bf k})|^2-\Omega_z^2({\bf k})) \rangle 
\propto k_{\rm F}^6 \propto n_{e}^2$. On the other hand, the electron-impurity
scattering decreases with density because it is proportional to $\sim
n_i V_{k_{\rm F}}^2 \propto n_{e}/k_{\rm F}^4 \propto n_{e}^{-1/3}$. Consequently, the
SRT decreases with density.

Therefore, both the electron-electron and electron-impurity
scatterings contribute to the nonmonotonic density dependence of the
SRT. The peak density $n_c$ appears in the crossover regime, where the
corresponding Fermi temperature is comparable with the lattice
temperature. A careful calculation gives the peak density as
$n_c=1.4\times10^{16}$~cm$^{-3}$, with the corresponding Fermi
temperature being 37~K, close to the lattice temperature
of 40~K.

We further discuss the electron density dependence of the SRT with
strain-induced SOC. In Fig.~\ref{fig:ne}(b), we plot the density
dependence of the SRT for $T=40$~K under strain with $n_i=n_{e}$. We
choose $\beta=2.6$~meV$\cdot$\AA.\cite{Kato} For this value, at low
density the SOC is dominated by the linear term due to strain.
However, in high density regime ($n_{e}>2\times 10^{17}$~cm$^{-3}$), the cubic
Dresselhaus term can surpass the linear term. It is noted that the
density dependence is also nonmonotonic and exhibits a peak. The
underlying physics is similar: In the non-degenerate regime, the SRT
increases with the density because both electron-electron and
electron-impurity scattering increase with density and the
inhomogeneous broadening changes little. In the degenerate regime, the
inhomogeneous broadening increases as $\propto k_{\rm F}^2\propto
n_{e}^{2/3}$, whereas both the electron-electron and electron-impurity
scatterings decrease with the electron density, which thus leads to
the decrease of the SRT with density.
It is further noted that the peak is at $n_c=2\times10^{16}$~cm$^{-3}$
which is larger than the peak density in the case without strain. 
This is because that the inhomogeneous broadening here increases as
$\langle k^2\rangle$, which is much slower than that of $\langle
k^6\rangle$ in the strain-free case. Nevertheless, the increase of the
scattering with density remains the same, therefore the peak shows up
at a larger electron density. This is also confirmed in the figure
that the SRT with only the linear term decreases slower than the one
with only the cubic Dresselhaus term in the high density
regime. However, in the low density regime, the SRT in the case with
strain increases as fast as the one in the strain-free case. This is
because here the increase of the SRT is due to the increase of the
scattering, whereas the inhomogeneous broadening changes little.

\subsubsection{Temperature dependence}

\begin{figure}[htb]
\includegraphics[height=5.cm]{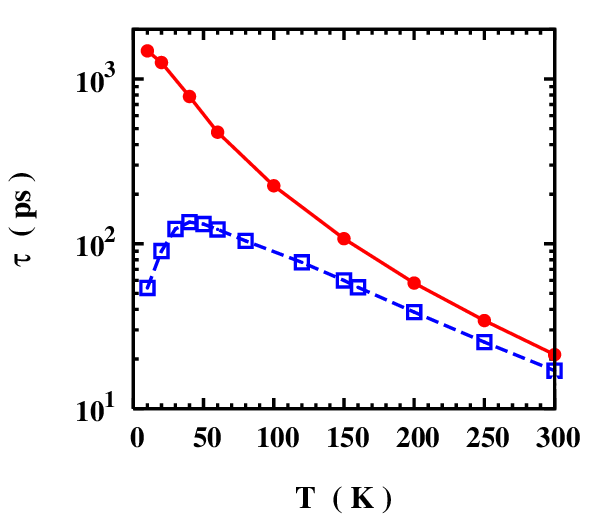}
\caption{(Color online) $n$-GaAs. SRT $\tau$ as function of $T$ for
  $n_i=n_{e}$ (solid curve with $\bullet$) and $n_i = 0.01n_{e}$ (dashed
  curve with $\square$). $n_{e}=10^{17}$~cm$^{-3}$.} 
\label{fig:T}
\end{figure}

We now study the temperature dependence of the SRT.
In Fig.~\ref{fig:T}, the SRT as function of temperature is
plotted for $n_{e}=10^{17}$~cm$^{-3}$. From the figure, it is seen
that the SRT decreases with temperature monotonically, which coincides
with previous experimental results.\cite{Awsch98,InAs,InAs2,GaAs1,GaN,Oertel}
This trend is also the same as that in the 2DES
with high impurity density.\cite{lowT} However, for high mobility
2DES, which can be achieved by modulation
doping, the temperature dependence of the SRT is nonmonotonic and
there is a peak $T_c$ around the Fermi temperature due to the
electron-electron scattering.\cite{lowT} 
In $n$-type bulk materials, as the impurity density is always equal to or
larger than the electron density, this peak disappears. Nevertheless,
similar effect can be obtained if one artificially reduces the
impurity density. For example when $n_i=0.01 n_{e}$, it is seen from
Fig.~\ref{fig:T} that the SRT shows nonmonotonic behavior with a peak
around 40~K.

\subsubsection{Initial spin polarization dependence: 
the effect of the Coulomb HF term}

We now turn to the dependence of initial spin polarization on spin
relaxation. Previously, it was discovered that at finite spin polarization,
the Coulomb HF term serves as an effective magnetic field along the
direction of the spin polarization.\cite{highP} The effective magnetic
field can be as large as 40~T at high spin polarization in
2DES which suppresses the DP spin
relaxation.\cite{highP} This effect was first predicted by Weng and
Wu\cite{highP} and then confirmed by experiments very
recently.\cite{wu-exp-hP,Zheng,tobias} However, the 
effect of the Coulomb HF term on spin relaxation in bulk system still
needs to be evaluated. Here, we present such an investigation.

\begin{figure}[htb]
\includegraphics[height=5.cm]{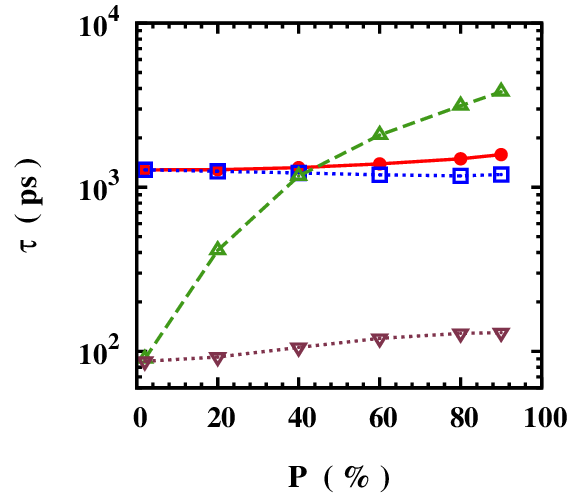}
\caption{(Color online) $n$-GaAs. Dependence of SRT $\tau$ on initial spin
  polarization $P$ for $n_i=n_{e}$ with (without) the Coulomb HF 
  term [curve with $\bullet$ ($\square$)] and for $n_i=0.01n_{e}$ with
  (without) the Coulomb HF term [curve with $\triangle$
  ($\triangledown$)]. $n_{e}=10^{17}$~cm$^{-3}$ and $T=20$~K.}
\label{fig:diffP1}
\end{figure}

The Coulomb HF term for spin polarization along, e.g.,
the $z$-direction can be written as
\begin{equation}
  \hat{\Sigma}_{\rm HF} ({\bf k}) = - \sum_{{\bf k}^{\prime}} V_{{\bf k}-{\bf
    k}^{\prime}} \left(f_{{\bf k}^{\prime}\uparrow}-f_{{\bf
    k}^{\prime}\downarrow}\right) \hat{s}_z.
\end{equation}
The corresponding effective magnetic field is along the $z$-axis,
\begin{equation}
  B_{\rm HF} ({\bf k}) = -\sum_{{\bf k}^{\prime}} V_{{\bf k}-{\bf
    k}^{\prime}} \left(f_{{\bf k}^{\prime}\uparrow}-f_{{\bf
    k}^{\prime}\downarrow}\right) \Big/g\mu_{\rm B}.
\label{HFB}
\end{equation}
Under this effective magnetic field, 
the spin precession is blocked and the SRT is
elongated.\cite{highP,Ivchenko} 
Recently, this effective magnetic field has been probed
experimentally,\cite{Zheng,tobias} and its effect on spin accumulation
in 2DES was also discussed theoretically.\cite{Bauer}

In Fig.~\ref{fig:diffP1}, we plot the SRT as function of the initial
spin polarization $P$ for $n_{e}=10^{17}$~cm$^{-3}$ and $T=20$~K.
It is seen that the SRT increases with the initial spin
polarization. To elucidate the effect of the Coulomb HF term, we also
plot the results from the calculation without the Coulomb HF term.
The results indicate that the increase of the
SRT with initial spin polarization is due to the Coulomb HF
term. However, the increment is less than
$50~\%$. Previously, it was shown in 2DES that the SRT can
increase over 30~times for low impurity case at 120~K, while less than
3~times for high impurity or high temperature case where the
scattering is strong.\cite{highP} The results can be understood as
follows: The SRT under the HF effective magnetic field can be
estimated as\cite{opt-or} 
\begin{equation}
  \tau_s(P) = \tau_s(P=0)[1+(g\mu_{\rm B} B_{\rm HF}\tau_p^{\ast})^2]
\label{HFeq}
\end{equation}
where $B_{\rm HF}$ is
the averaged effective magnetic field. Thus the effect of the HF
effective magnetic field increases with $\tau_p^{\ast}$, i.e., the effect is
more pronounced for weak scattering case, such as the low impurity
density case. However, in bulk system the
impurity density is always equal to or larger than the electron
density $n_i\ge n_{e}$, therefore the effect of the Coulomb HF term is
suppressed. This can be seen from the calculation with $n_i=0.01n_{e}$
(the artificial case). The results are also plotted in the figure.
One finds that the Coulomb HF term effectively enhances the SRT
at low impurity density by 40~times. Therefore, due to the large impurity density
($n_i\ge n_{e}$), the SRT is insensitive to the initial spin
polarization in $n$-type bulk III-V semiconductors. We have checked
that the conclusion holds for other cases with different temperatures,
electron densities and materials.

\section{spin relaxation in intrinsic III-V semiconductors}

\begin{figure}[htb]
\includegraphics[height=5.cm]{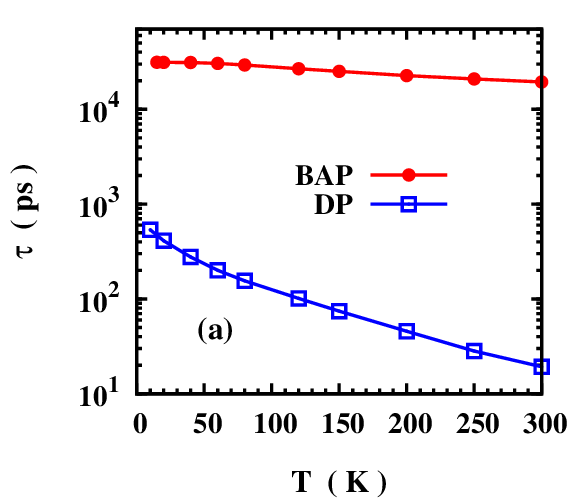}
\includegraphics[height=5.cm]{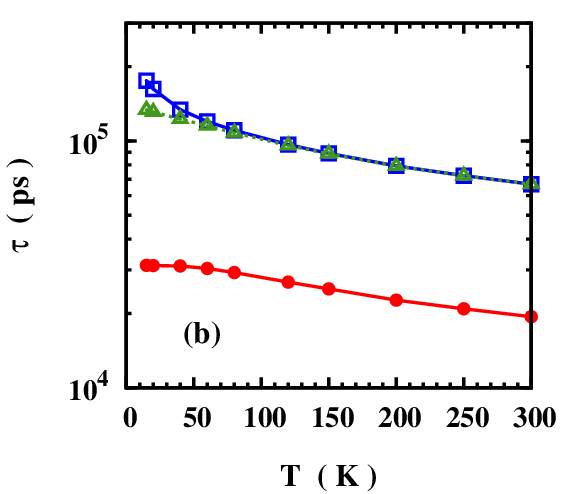}
\includegraphics[height=5.cm]{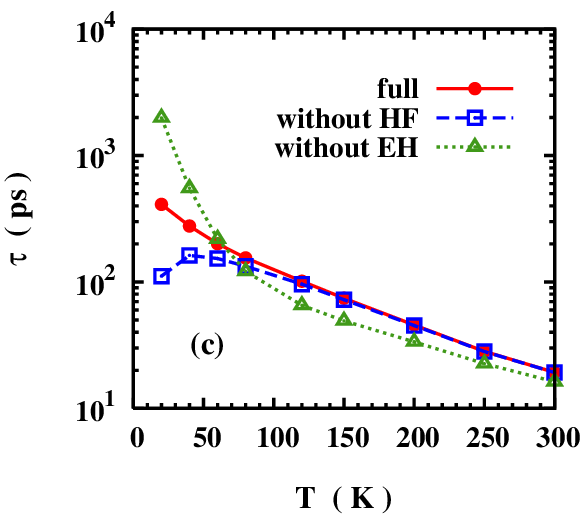}
\caption{(Color online) Intrinsic GaAs with
  $N_{\rm ex}=10^{17}$~cm$^{-3}$. (a) SRT $\tau$ due to the BAP and DP
  mechanisms as function of temperature. (b) SRT due to the BAP
  mechanism calculated from Eq.~(\ref{BAPapp1}) (dotted curve with 
  $\triangle$), from the KSBE approach with both long-range
  and short-range exchange scatterings (solid curve with $\bullet$) as
  well as from the KSBE approach with only the short-range exchange scattering (solid
  curve with $\square$). (c) SRT due to the DP mechanism from full
  calculation (solid curve with $\bullet$), from the calculation
  without the Coulomb HF term (dashed curve with $\square$), and from
  the calculation without the electron-hole Coulomb scattering (dotted
  curve with $\triangle$). For electrons $T_{\rm F}=136$~K, and for holes $T_{\rm F}=16$~K.}
\label{fig:T_int}
\end{figure}

In this section we study spin relaxation in intrinsic III-V
semiconductors. For intrinsic semiconductors, the electrons and holes are
created by optical excitation, and their numbers are equal. Compared to
$n$-type semiconductors, there are two new scattering mechanisms: the
electron-hole Coulomb and electron-hole exchange scatterings, where the
latter corresponds to the BAP mechanism. Another
important property of intrinsic semiconductors is that the impurity
density is very low (we take $n_i=0$), which offers a good 
platform for
demonstrating the effect of the many-body carrier-carrier scattering on
spin relaxation. Moreover, the effect of the Coulomb HF term would be
enhanced as the electron-impurity scattering can be eliminated. Our
first goal is to compare the relative efficiency of the DP and BAP
mechanisms. After that we also study the temperature and
photo-excitation density $N_{\rm ex}$ dependences of the SRT. The role of
electron-hole Coulomb scattering as well as the effect of the Coulomb HF term
are also addressed. We further compare the results from the KSBEs with
those from the widely used analytical formulae
[Eqs.~(\ref{BAPapp1}) and (\ref{BAPapp2})]. The initial spin 
polarization is chosen to be $50~\%$ which corresponds to circularly
polarized optical excitation. We focus on GaAs, while the situation is
similar for other III-V semiconductors.\cite{note5}

\subsection{Temperature dependence}

In Fig.~\ref{fig:T_int}(a), we plot the SRTs due to the DP and BAP
mechanisms as function of temperature for
$N_{\rm ex}=10^{17}$~cm$^{-3}$. The SRT due to the BAP mechanism alone is
calculated by removing the spin precession due to the SOC, but keeping
all the scattering terms. It is noted that the SRT due to the BAP
mechanism is larger than that due to the DP mechanism by more than one
order of magnitude, which indicates that the BAP mechanism is
negligible for intrinsic GaAs.\cite{note7} Moreover, the spin relaxation due to
the DP mechanism increases with temperature more rapidly than that due to
the BAP mechanism at high temperature. This is because the increase of
spin relaxation due to the DP mechanism mainly comes from the increase
of the inhomogeneous broadening which is proportional to $T^3$ in high
temperature (non-degenerate) regime. Meanwhile, according to
Eq.~(\ref{BAPapp1}), the increase of spin relaxation due to the BAP
mechanism mainly comes from the increase of $\langle v_k\rangle$ which
is proportional to $T^{0.5}$ in that regime.

For a close examination of the BAP mechanism, we also plot the SRT
limited by the the short-range electron-hole exchange scattering
calculated from the KSBEs in Fig.~\ref{fig:T_int}(b). It is seen that the
SRT limited by the short-range electron-hole exchange scattering is
much larger than that limited by both long- and short-range exchange
scattering. This confirms that the long-range scattering is more
important than the short-range one in GaAs as $\Delta E_{\rm LT}$ is four
times larger than $\Delta E_{\rm SR}$. Therefore, previous
investigations\cite{Song,Fishman,spintronics} with only
the short-range exchange scattering included are
questionable. Moreover, to check the validity of the widely used
elastic scattering approximation, we also compare the results from the
KSBEs with those from the elastic scattering approximation.
Under the elastic scattering approximation,\cite{opt-or}
\begin{eqnarray}
\frac{1}{\tau_{\rm BAP}({\bf k})} &=& 4\pi \sum_{{\bf q},{\bf k}^{\prime},m,m^{\prime}} 
\delta(\varepsilon_{{\bf k}} +\varepsilon^{h}_{{\bf
    k}^{\prime} m^{\prime}}-\varepsilon_{{\bf k}-{\bf q}} -
\varepsilon^{h}_{{\bf k}^{\prime}+{\bf q}m}) \nonumber \\ 
&&\mbox{} \times |{\cal J}^{(-)\ {\bf k}^{\prime}+{\bf q}m}_{{\bf k}^{\prime}
  m^{\prime}}|^2 f^{h}_{{\bf k}^{\prime} m^{\prime}} 
      \left(1 - f^{h}_{{\bf k}^{\prime}+{\bf q}m}\right).
\label{BAPelastapp1}
\end{eqnarray}
The SRT is then obtained by averaging over the electron distribution.
Eqs.~(\ref{BAPapp1}) and (\ref{BAPapp2}) are derived from the above
equation under some approximations. For example, by including only the
short-range exchange scattering and ignoring the light-hole 
contribution as well as the term of
$(1 - f^{h}_{{\bf k}^{\prime}+{\bf q}m})$, Eq.~(\ref{BAPapp1}) is
obtained. To show that the elastic scattering approximation fails in the
degenerate regime, we compare our results with the results from 
Eq.~(\ref{BAPelastapp1}). For simplicity, we include only the
short-range exchange scattering. In Fig.~\ref{fig:T_int}(b), we plot
the SRT obtained from Eq.~(\ref{BAPelastapp1}) as the dotted curve. It
is seen that the result from Eq.~(\ref{BAPelastapp1}) agrees well with
our result from the KSBEs at high temperature, but deviates at low
temperature. The deviation is due to the Pauli blocking of electrons
in the degenerate regime, which is neglected in the elastic scattering
approximation ($T_{\rm F}=136$~K).\cite{note6}

\begin{figure}[htb]
\includegraphics[height=5.5cm]{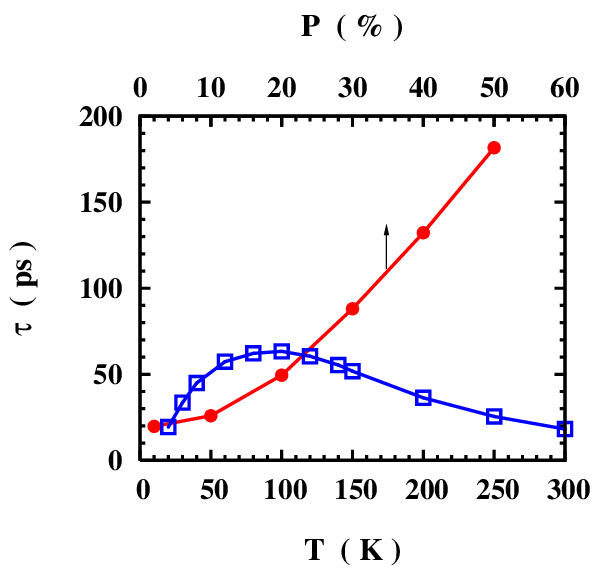}
\caption{(Color online) Intrinsic GaAs with $N_{\rm ex}=2\times
  10^{17}$~cm$^{-3}$. SRT $\tau$ as function of temperature for
  $P=2~\%$ (curve with $\square$) and the SRT as function of initial
  spin polarization $P$ for $T=20$~K (curve with $\bullet$) (note that
  the scale of $P$ is on the top of the frame).}
\label{fig:TP_int}
\end{figure}

We also discuss the effect of the electron-hole Coulomb scattering and
the Coulomb HF term on spin relaxation due to the DP mechanism. In
Fig.~\ref{fig:T_int}(c), we plot the SRTs due to the DP mechanism
obtained from the full calculation, from the calculation without
the electron-hole Coulomb scattering, and from the calculation without the
Coulomb HF term. Let us first examine the effect of the Coulomb HF
term on spin relaxation. It is seen that the Coulomb HF term has
important effect on spin relaxation only for low temperature
case\cite{HFtaup} ($T<60$~K) which is consistent with the results in
2DES.\cite{highP} We then turn to the effect of the electron-hole Coulomb
scattering. It is seen that without the electron-hole Coulomb scattering the
SRT is larger for $T<60$~K but smaller for $T> 60$~K compared
with that from the full calculation. This behavior can be understood as
following: For $T< 60$~K, the Coulomb HF term has important effect on
spin relaxation. The HF effective magnetic field elongates the
SRT. According to Eq.~(\ref{HFeq}), this effect increases
with the momentum scattering time. Without the electron-hole Coulomb scattering
the momentum scattering time is elongated, which enhances the effect
and leads to longer SRT. For higher temperature ($T>60$~K), the effect
of the Coulomb HF term is weak,\cite{HFtaup} and the system
returns back to the motional narrowing regime. The SRT thus decreases
when the electron-hole Coulomb scattering is removed. The results indicate
that the electron-hole Coulomb scattering is comparable with the
electron-electron and electron--LO-phonon scatterings. In other words,
besides the screening from holes, the main contribution of the hole
system to electron spin relaxation comes from the electron-hole Coulomb
scattering in intrinsic semiconductors.

In Fig.~\ref{fig:TP_int}, we plot the SRT as function of temperature
for $N_{\rm ex}=2\times 10^{17}$~cm$^{-3}$ with $P=2~\%$. 
In Sec.~III~B2, we showed that there is a peak in the temperature
dependence of SRT due to the electron-electron Coulomb scattering when the
impurity density is low. In intrinsic semiconductors, as the impurity density is
very low, the peak may appear. Indeed, we find that the SRT
has a peak at $T_c\sim 100$~K. The peak temperature
$T_c$ is comparable with the Fermi temperature ($T_{\rm F}=216$~K)
[Actually, our calculation indicates that the peak temperature
 $T_c$ is around $T_{\rm F}/3$ and lies in the range of $(T_{\rm F}/4, T_{\rm F}/2)$
depending on  the carrier density.]
Nevertheless, at high spin polarization, such as $P=50~\%$,
the peak disappears as indicated in Fig.~\ref{fig:T_int}(c). This peak
can be observed within current technology of optical
orientation. However, up till now, no such experimental investigation
has been performed. In Fig.~\ref{fig:TP_int}, we also plot the SRT as
function of initial spin polarization at $T=20$~K. It is seen that the
SRT is elongated by 9~times when $P$ is tuned from $2~\%$ to
$50~\%$. Therefore, the effect of the Coulomb HF term can also be
observed in intrinsic materials and is more pronounced compared to the
$n$-type case.

\begin{figure}[htb]
\includegraphics[height=5.cm]{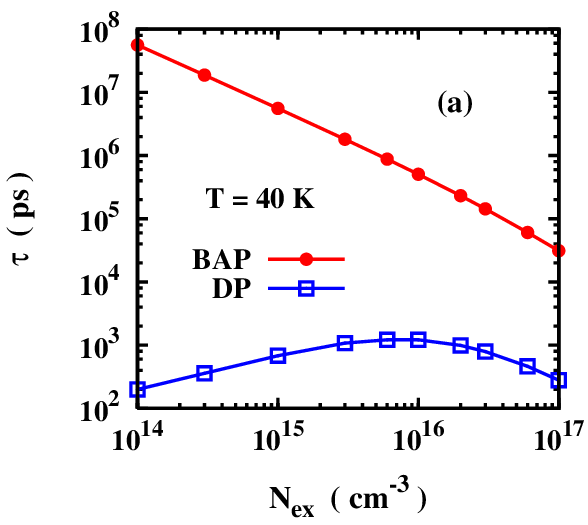}
\includegraphics[height=5.cm]{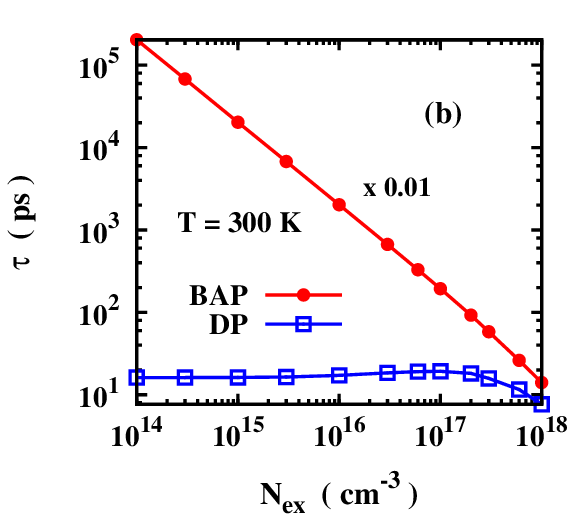}
\includegraphics[height=5.cm]{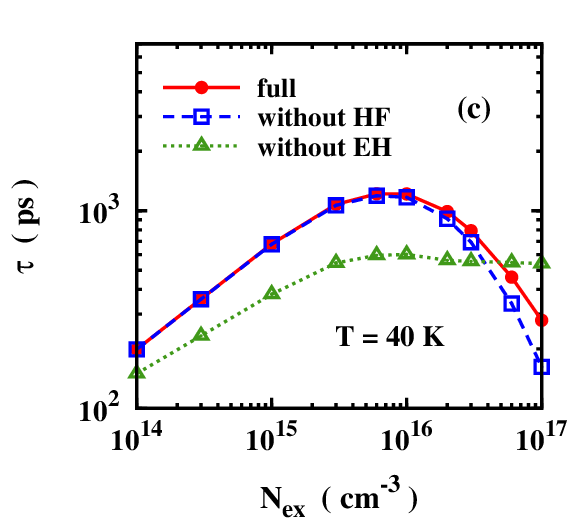}
\caption{(Color online) Intrinsic GaAs. SRT $\tau$ due to the BAP
  mechanism and that due to the DP mechanism as function of photo-excitation
  density $N_{\rm ex}$ at: $T=40$~K (a) and $T=300$~K (b) (note that the
  value of $\tau_{\rm BAP}$ in figure (b) has been rescaled by a
  factor of 0.01). (c): the SRT due
  to the DP mechanism from the full calculation (curve with $\bullet$),
  from the calculation without the Coulomb HF term (curve with
  $\square$), and from the calculation without the electron-hole
  Coulomb scattering (curve with $\triangle$) for $T=40$~K.}
\label{fig:ne_int}
\end{figure}

\subsection{Density dependence}

We plot the density dependence of the SRT in
Fig.~\ref{fig:ne_int} for both low temperature ($T=40$~K) and room
temperature ($T=300$~K) cases. It is seen that for both cases
the BAP mechanism is far less efficient than the DP mechanism. Another
remarkable feature is that the SRT shows a {\em nonmonotonic}
photo-excitation density dependence with a peak at some density $n_c$
which resembles that in $n$-type materials.
Further calculation gives $n_c=0.8\times
10^{16}$~cm$^{-3}$ ($T_{\rm F}=25$~K) for $T=40$~K case and $n_c=0.9\times
10^{17}$~cm$^{-3}$ ($T_{\rm F}=127$~K) for $T=300$~K case.
Interestingly, two recent experiments give different photo-excitation
density dependences of SRT at room temperature: in
Ref.~\onlinecite{Lai} the SRT decreases with $N_{\rm ex}$ where
$N_{\rm ex}>10^{17}$~cm$^{-3}$, while in Ref.~\onlinecite{Oertel} the SRT
increases with $N_{\rm ex}$ where the photo-excitation density is lower. These
observations are consistent with our results. However, the peak has
not been reported in the literature.

We then discuss the effects of the electron-hole Coulomb scattering and the
Coulomb HF term on the spin relaxation for $T=40$~K
as function of photo-excitation density. In Fig.~\ref{fig:ne_int}(c), we plot
the SRTs obtained from the full calculation, from
the calculation without the electron-hole Coulomb scattering, and from the
calculation without the Coulomb HF term. It is seen that the Coulomb
HF term plays a visible role only for high densities, as the HF
effective magnetic field increases with electron density.\cite{HFtaup}
Similar to the temperature dependence [Fig.~\ref{fig:T_int}(c)],
without the electron-hole Coulomb scattering, the SRT is larger for $N_{\rm ex} >
3\times 10^{16}$~cm$^{-3}$ where the Coulomb HF term plays a prominent
role, while it is smaller for lower photo-excitation densities where the
Coulomb HF term is unimportant. Notably, the peak of SRT still exists
when the Coulomb HF term is removed, which implies that the degree of
initial spin polarization is irrelevant for the existence of the peak.

\section{electron spin relaxation in $p$-type III-V semiconductors}

In this section, we study spin relaxation in $p$-type III-V
semiconductors. The main sources of spin relaxation
have been recognized as the BAP mechanism and the DP
mechanism.\cite{note3} We first compare the relative efficiency
of the two mechanisms for various hole densities and
temperatures. After that, the hole density and the photo-excitation
density dependences of the SRT at given temperature are also
discussed.

\subsection{Comparison of the DP and BAP mechanisms in GaAs}

We first address the relative importance of the BAP and DP mechanisms
for various hole densities and temperatures in GaAs. The electrons
are created by photo-excitation with $P=50~\%$ (i.e., we assume ideal
optical orientation by circularly polarized light).\cite{note8} In order to avoid
exaggerating the DP mechanism, we use the SOC parameter fitted from
the experimental data in Ref.~\onlinecite{Awsch98}, i.e.,
$\gamma_{\rm D}=8.2$~eV$\cdot$\AA$^3$ (see Appendix~B) throughout this
section, which is smaller than the value from the 
${\bf k}\cdot{\bf p}$ calculation
$\gamma_{\rm D}=23.9$~eV$\cdot$\AA$^3$.\cite{wu-soc,soc35}

\subsubsection{Low photo-excitation}

\begin{figure}[htb]
\includegraphics[height=5cm]{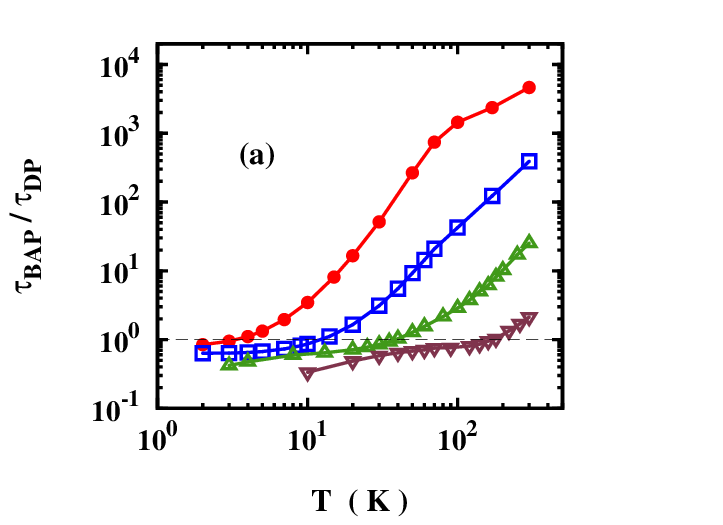}
\includegraphics[height=5cm]{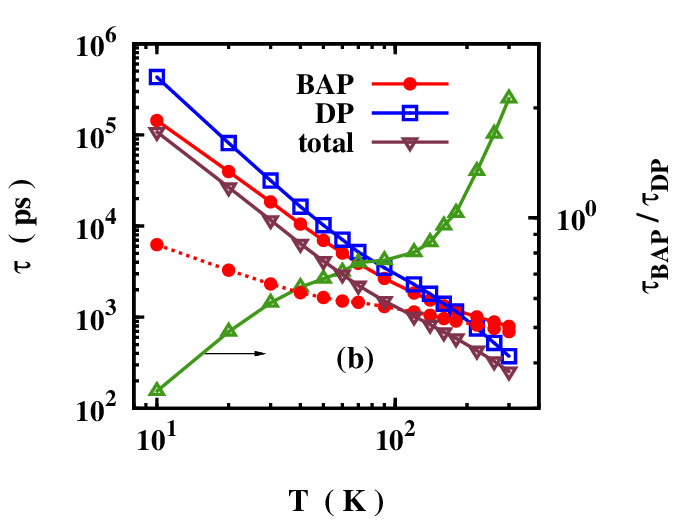}
\caption{(Color online) $p$-GaAs. Ratio of the SRT due to the BAP
  mechanism to that due to the DP mechanism as function of temperature
  for various hole densities with $N_{\rm ex}=10^{14}$~cm$^{-3}$ and
  $n_i=n_h$. (a): $n_h=3\times 10^{15}$~cm$^{-3}$ (curve with $\bullet$), $3\times
  10^{16}$~cm$^{-3}$ (curve with $\square$),
  $3\times 10^{17}$~cm$^{-3}$ (curve with $\triangle$), and 
  $3\times 10^{18}$~cm$^{-3}$ (curve with $\triangledown$).
  The hole Fermi temperatures for these densities are $T_{\rm F}^{h}=1.6$, 7.3,
  34, and 156~K, respectively. The electron Fermi temperature is $T_{\rm F}=1.4$~K.
  (b): The SRTs due to the BAP and DP mechanisms, the total SRT,
  together with the ratio $\tau_{\rm BAP}/\tau_{\rm DP}$ (curve with
  $\triangle$) versus the temperature for $n_h=3\times
  10^{18}$~cm$^{-3}$. The dotted curve represents the SRT due to the
  BAP mechanism without the Pauli blocking of holes. Note the scale of
  $\tau_{\rm BAP}/\tau_{\rm DP}$ is on the right hand side of the
  frame.}
\label{fig:T_p}
\end{figure}

We first concentrate on low photo-excitation density regime, where
we choose $N_{\rm ex}=10^{14}$~cm$^{-3}$. The ratio of the SRT due to the
BAP mechanism to that due to the DP mechanism is plotted in
Fig.~\ref{fig:T_p}(a) for various hole densities. It is seen that 
the DP mechanism dominates at high temperature, whereas the BAP
mechanism dominates at low temperature, which is consistent with
previous investigations.\cite{Aronov,Song,Fishman,spintronics,opt-or}
An interesting feature is that the ratio first decreases rapidly, then
slowly, and then again rapidly with decreasing temperature. A typical
case is shown in Fig.~\ref{fig:T_p}(b) for $n_h=3\times
10^{18}$~cm$^{-3}$. It is noted that the ``plateau'' is around the
hole Fermi temperature $T_{\rm F}^{h}=156$~K which is given by 
\begin{equation}
  T_{\rm F}^{h}=\frac{(3\pi^2 n_h)^{2/3}}{2 k_{\rm B} m_0\left[\left(\gamma_1-2\gamma_2\right)^{-3/2} + \left(\gamma_1+2\gamma_2\right)^{-3/2}\right]^{2/3}}.
\end{equation}
The underlying physics is that: on one hand, the Pauli blocking of
holes becomes important when $T\lesssim T_{\rm F}^{h}$,
which slows down the BAP spin relaxation effectively [see Eq.~(\ref{BAPapp2})]; on
the other hand, the increase of the screening (mainly from holes) with
decreasing temperature weakens the electron-impurity and
carrier-carrier  scatterings and thus enhances the DP spin  
relaxation. Consequently, the decrease of the ratio with decreasing
temperature slows down and the ``plateau'' is formed around $T_{\rm F}^{h}$. However, after
the hole system enters  the degenerate regime, the screening
changes little with temperature. The ratio thus decreases rapidly with
decreasing temperature again [see Appendix~C]. We also plot the SRT due to the BAP mechanism
without the Pauli blocking of holes as dotted curve in Fig.~\ref{fig:T_p}(b),
which indicates that the Pauli blocking of holes effectively suppresses the BAP
spin relaxation at low temperature ($T\lesssim T_{\rm F}^{h}$). It is also seen
from Fig.~\ref{fig:T_p}(b) that the total SRT increases with decreasing
temperature as both $\tau_{\rm DP}$ and $\tau_{\rm BAP}$ do.

\subsubsection{High photo-excitation}

\begin{figure}[htb]
\includegraphics[height=5cm]{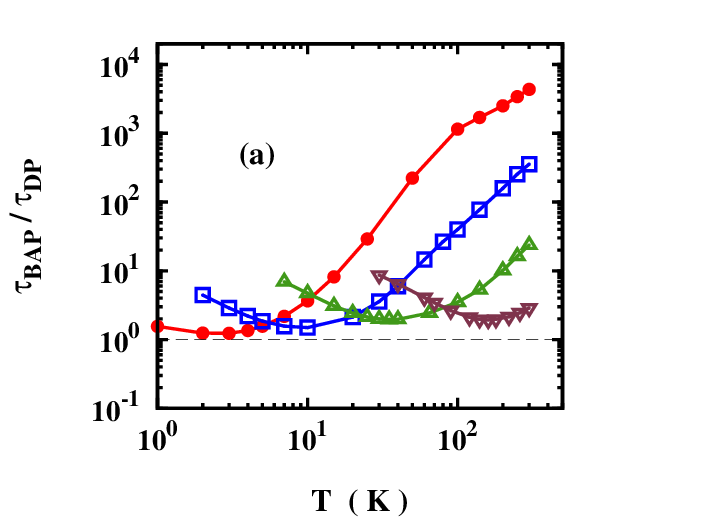}
\includegraphics[height=5cm]{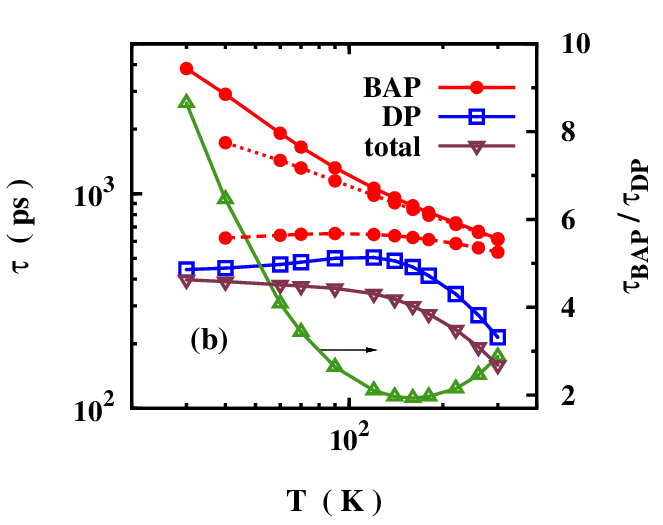}
\caption{(Color online) $p$-GaAs. Ratio of the SRT due to the BAP
  mechanism to that due to the DP mechanism as function of temperature
  for various hole densities with $N_{\rm ex}=0.1n_h$ and
  $n_i=n_h$. (a): $n_h=3\times 10^{15}$~cm$^{-3}$ (curve with $\bullet$), $3\times
  10^{16}$~cm$^{-3}$ (curve with $\square$), $3\times 10^{17}$~cm$^{-3}$
  (curve with $\triangle$), and  $3\times 10^{18}$~cm$^{-3}$ (curve
  with $\triangledown$). The hole Fermi temperatures for these
  densities are $T_{\rm F}^{h}=1.7$, 7.7, 36, and 167~K, respectively. The electron Fermi
  temperatures are $T_{\rm F}=2.8$, 13, 61, and 283~K, respectively.
  (b): The SRTs due to the BAP and DP mechanisms, the total SRT,
  together with the ratio $\tau_{\rm BAP}/\tau_{\rm DP}$ (curve with
  $\triangle$) versus the  temperature for $n_h=3\times
  10^{18}$~cm$^{-3}$. The dotted (dashed) curve represents the SRT due
  to the BAP mechanism without the Pauli blocking of electrons (holes). 
  Note the scale of $\tau_{\rm BAP}/\tau_{\rm DP}$
  is on the right hand side of the frame.}
\label{fig:T_p2}
\end{figure}

We then discuss the case with high photo-excitation density, where
we choose $N_{\rm ex}=0.1n_h$. The ratio of the SRT due to the BAP
mechanism to that due to the DP mechanism is plotted in
Fig.~\ref{fig:T_p2}(a) for various hole densities. It is seen that, 
interestingly, the ratio is nonmonotonic and has a minimum roughly
around the Fermi temperature of electrons, $T\sim T_{\rm F}$. The BAP
mechanism is comparable with the DP mechanism only in the moderate
temperature regime roughly around $T_{\rm F}$, whereas for higher or lower
temperature it becomes unimportant. To explore the underlying physics,
we plot the SRTs due to the BAP and DP mechanisms in
Fig.~\ref{fig:T_p2}(b) for $n_h=3\times 10^{18}$~cm$^{-3}$. It
is seen that the Pauli blocking of electrons and holes largely
suppresses the BAP spin relaxation in the low temperature regime
and hence makes $\tau_{\rm BAP}$ always increase with decreasing
temperature. On the other hand, $\tau_{\rm DP}$ first increases with
decreasing temperature, then saturates at low temperature ($T<T_{\rm F}$) 
due to the fact that both the inhomogeneous broadening and the momentum
scattering change little in the degenerate regime. Therefore, the
ratio $\tau_{\rm BAP}/\tau_{\rm DP}$ first decreases then increases
with decreasing temperature and shows a minimum roughly around the
electron Fermi temperature. This scenario holds for arbitrary
excitation density, and the temperature where the ratio 
$\tau_{\rm BAP}/\tau_{\rm DP}$ reaches its minimum increases with
excitation density. Finally, it is seen that the total SRT saturates
at low temperature as $\tau_{\rm DP}$ does.

\begin{figure}[htb]
\includegraphics[height=5.cm]{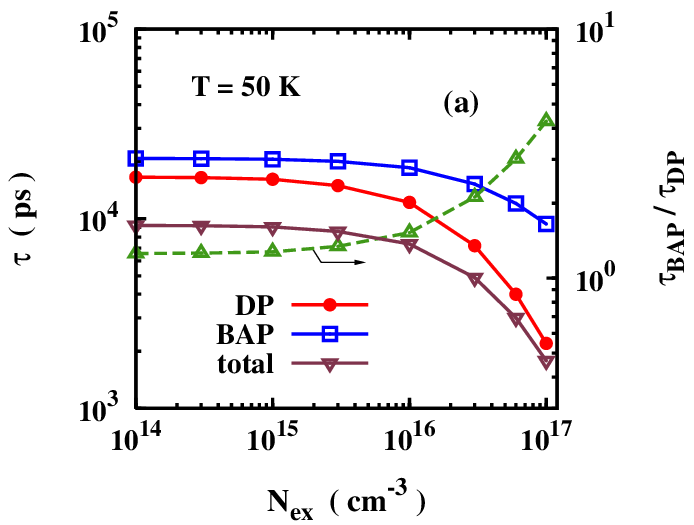}
\includegraphics[height=5.cm]{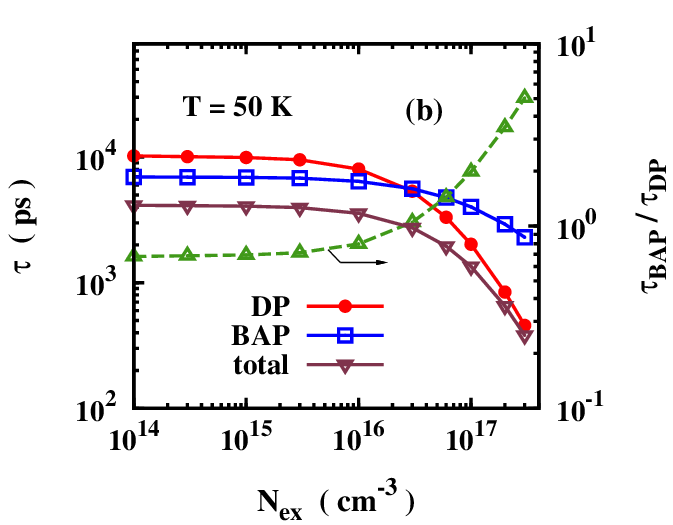}
\caption{(Color online) $p$-GaAs. SRTs $\tau$ due to the BAP and DP
  mechanisms together with the total SRT versus the photo-excitation density 
  $N_{\rm ex}$. The ratio of the two is plotted as dashed curve (note that
  the scale is on the right hand side of the frame). (a): $n_i=n_h=3\times
  10^{17}$~cm$^{-3}$. (b): $n_i=n_h=3\times 10^{18}$~cm$^{-3}$. $T=50$~K.}
\label{fig:nex_p}
\end{figure}

\subsection{Photo-excitation density dependence}

We now turn to the photo-excitation density dependence of the SRT. In
Fig.~\ref{fig:nex_p}(a), we plot the SRT due to the DP mechanism and that due
to the BAP mechanism as function of photo-excitation density for 
$n_h=3\times 10^{17}$~cm$^{-3}$ with $T=50$~K. It is seen that the
SRT due to the DP mechanism decreases with the photo-excitation density
monotonically. Specifically, it first decreases slowly, then
($N_{\rm ex}>10^{16}$~cm$^{-3}$) rapidly with the photo-excitation
density. The scenario is as follows: In $p$-type semiconductors at
low temperature, the dominant scattering mechanisms are the electron-hole
and electron-impurity scatterings. The momentum
scattering due to these two mechanism changes little with
photo-excitation (electron) density ($N_{\rm ex}=n_{e}$) for $N_{\rm ex}<n_h$. In the low
density regime ($n_{e}<3\times 10^{15}$~cm$^{-3}$, or
$T_{\rm F}<13$~K), where the electron system is non-degenerate, the increase
of density affects the inhomogeneous broadening very little. Thus
the SRT changes slowly with the photo-excitation density. In the high
density regime ($n_{e}>3\times 10^{16}$~cm$^{-3}$, or $T_{\rm F}>61$~K),
the electron system is degenerate where the inhomogeneous broadening
increases fast with density. Consequently the SRT decreases rapidly
with the photo-excitation density. For the SRT due to the BAP mechanism,
it decreases slowly with the photo-excitation density in the low density
regime, but rapidly in the high density regime. The
decrease is mainly due to the increase of the averaged
electron velocity $\langle v_{k}\rangle$ [see Eq.~(\ref{BAPapp2})],
which is determined by the temperature and is insensitive to density in
the non-degenerate regime, but increases rapidly in the degenerate
regime. However, the increase of the spin
relaxation due to the BAP mechanism is slower than that due to the DP
mechanism, because the inhomogeneous broadening increases as
$\propto N_{\rm ex}^2$ while $\langle v_{k}\rangle$ increases as
$\propto N_{\rm ex}^{1/3}$. Consequently, the BAP mechanism becomes even less
important in the high photo-excitation density regime.
Similar situation also happens for other hole densities and
temperatures. In Fig.~\ref{fig:nex_p}(b), we plot the case for a
larger hole density $n_h=3\times 10^{18}$~cm$^{-3}$. It is
seen that under low photo-excitation, the BAP mechanism is more important
than the DP mechanism. However, the BAP mechanism becomes less
important than the DP mechanism in the high photo-excitation density
regime. The crossover of the low photo-excitation density regime to the
high photo-excitation density regime takes place around $T_{\rm F}\sim T$.
This leads to the conclusion that the BAP mechanism is not important
at high photo-excitation density in $p$-type materials. It
is seen from Fig.~\ref{fig:nex_p} that the total SRT decreases with
photo-excitation density as both $\tau_{\rm DP}$ and $\tau_{\rm BAP}$
do. This behavior is also consistent with what observed in
experiments in Ref.~\onlinecite{Nex-decr}.

\begin{figure}[htb]
\includegraphics[height=5.cm]{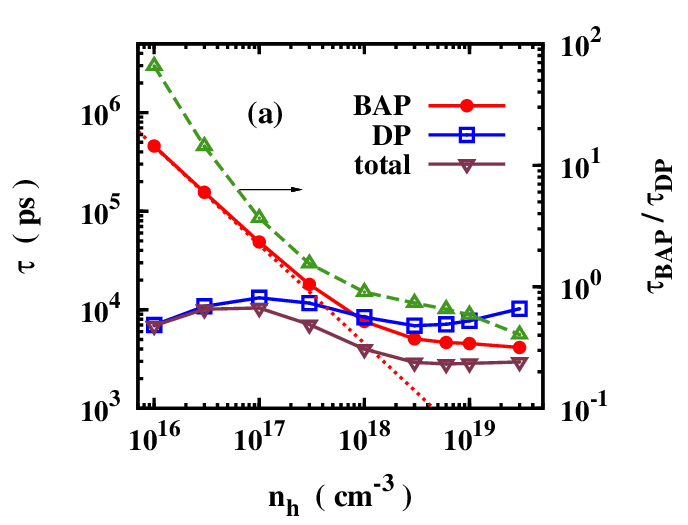}
\includegraphics[height=5.cm]{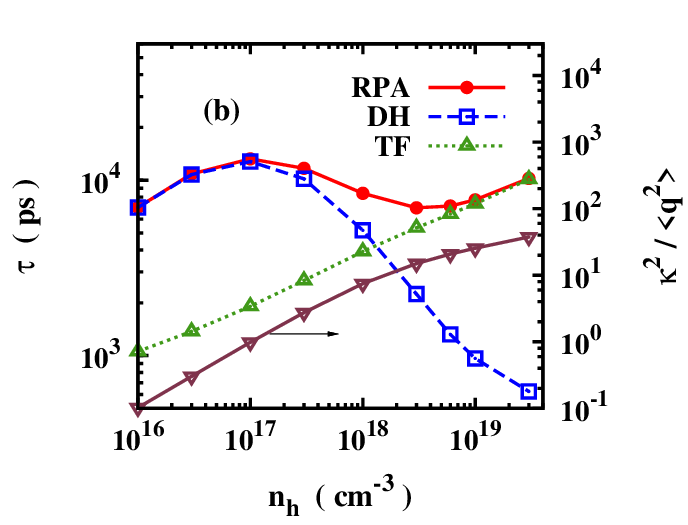}
\caption{(Color online) $p$-GaAs. (a): SRTs $\tau$ due to the BAP and
  DP mechanisms together with the total SRT against hole density
  $n_h$. $N_{\rm ex}=10^{14}$~cm$^{-3}$, $n_i=n_h$, and $T=60$~K.
  The dotted curve denotes a fitting of the curve with $\bullet$
  using $1/n_h$ scale. The curve with $\triangle$ denotes the ratio
  $\tau_{\rm BAP}/\tau_{\rm DP}$ (note that the scale is on the right
  hand side of the frame). (b): SRTs due to the DP mechanism with the
  DH (curve with $\square$), TF (curve with $\triangle$), and the RPA
  (curve with $\bullet$) screenings. The ratio $\kappa^2/\langle
  q^2\rangle$ is plotted as curve with $\triangledown$ (note that the
  scale is on the right hand side of the frame).} 
\label{fig:nh_p}
\end{figure}

\subsection{Hole density dependence}

We also study the hole density dependence of spin relaxation due
to the BAP and DP mechanisms. In Fig.~\ref{fig:nh_p}(a), we plot the SRT due
to the BAP mechanism and that due to the DP mechanism as function of hole
density for $T=60$~K and $N_{\rm ex}=10^{14}$~cm$^{-3}$. It is seen that the
SRT due to the BAP mechanism decreases as $1/n_h$ at low hole
density, which is consistent with Eq.~(\ref{BAPapp1}), i.e., 
for non-degenerate holes $\tau_{\rm BAP}\propto 1/n_h$. 
At high hole density, $\tau_{\rm BAP}$ decreases slower
than $1/n_h$ due to the Pauli blocking of holes.
However, for the SRT due to the DP mechanism, the dependence is not so
obvious: the SRT first increases, then decreases and again increases
with the hole density. As the electron distribution, and hence the
inhomogeneous broadening, does not change with the hole density, the
variation of the SRT due to the DP mechanism solely comes from the
momentum scattering (mainly from the electron-impurity scattering).
To elucidate the underlying physics, we plot the SRT due to the DP
mechanism calculated with the RPA screening together with those calculated
with the Thomas-Fermi (TF) screening\cite{koch} [which applies in the
  degenerate (high density) regime], the Debye-Huckle (DH)
screening\cite{koch} [which applies in the non-degenerate (low
  density) regime] in Fig.~\ref{fig:nh_p}(b). From the figure it
is seen that the first increase and the decrease is connected with the DH
screening, whereas the second increase is connected with the TF
screening. The underlying physics is as follows: In the low hole density
regime, the screening from the holes is small and the Coulomb
potential, which is proportional to $1/(\kappa^2+q^2)$, changes slowly
with the screening constant $\kappa$. Hence the
electron-impurity scattering increases with $n_h$ as it is
proportional to $n_i \langle V_q^2\rangle \propto n_h$ (as
$n_i=n_h$). For higher hole density ($n_h>10^{17}$~cm$^{-3}$), the 
screening constant $\kappa$ becomes larger than the transfered
momentum $q$. [To elucidate the relative ratio of the two, we plot the
ratio of the average of the square of the transfered momentum $\langle
q^2\rangle$ to the square of the screening constant $\kappa^2$ as
curve with $\triangledown$ in Fig.~\ref{fig:nh_p}(b).] Hence the
electron-impurity scattering decreases with $n_h$ because it is
proportional to $n_i \langle V_q^2 \rangle \propto n_h/\kappa^4\propto n_h^{-1}$ as
$\kappa^2 \propto n_h$ for the DH screening. As the hole density
increases, the hole system enters into the degenerate regime, where
the TF screening applies and $\kappa^2\propto n_h^{1/3}$. Hence, the
electron-impurity scattering increases with the hole density as $n_i
\langle V_q^2 \rangle \propto n_h^{1/3}$. Consequently, the SRT first
increases, then decreases and again increases with the hole density as
the momentum scattering does. It should be mentioned that this
behavior is different from that in the $p$-type (001) quantum wells
where $\tau_{\rm DP}$ increases with $n_i$ monotonically\cite{GaMnAs}
as the screening from holes is much weaker in that case due to
lower-dimension in phase-space and smaller hole effective mass
[in (001) GaAs quantum wells, the in-plane effective mass of
the  heavy-hole is $\sim 0.11 m_0$ compared to $0.54 m_0$ in bulk].

It is also noted in Fig.~\ref{fig:nh_p}(a) that the ratio $\tau_{\rm
  BAP}/\tau_{\rm DP}$ first decreases rapidly, then slowly and again
rapidly with the hole density $n_h$. The first decrease is because
that $\tau_{\rm BAP}$ decreases with $n_h$, whereas $\tau_{\rm DP}$
increases with it. In the crossover regime ($n_h\sim 10^{18}$~cm$^{-3}$),
where $T_{\rm F}\sim T$, the SRT due to the DP mechanism varies slowly
with hole density. As the SRT due to the BAP mechanism also
varies slowly with hole density due to the Pauli blocking of holes,
the ratio $\tau_{\rm BAP}/\tau_{\rm DP}$ changes slowly with hole
density in this regime and a ``plateau'' is formed at $T_{\rm F}\sim T$. In
higher hole density regime, however, $\tau_{\rm DP}$ increases with
$n_h$, whereas $\tau_{\rm BAP}$ decreases with it. The ratio
$\tau_{\rm BAP}/\tau_{\rm DP}$ hence decreases rapidly with $n_h$
again and the BAP mechanism becomes more and more important.
It is seen from Fig.~\ref{fig:nh_p}(a) that the
total SRT first increases then decreases in the low hole density
regime as $\tau_{\rm DP}$ does. Consequently, 
 the hole density dependence of the SRT exhibits a peak which has never been
reported. Nevertheless, in the regime of 
higher hole density, the total SRT changes slowly with $n_h$ 
as the BAP and DP mechanisms compete with each other.

\begin{figure}[htb]
\includegraphics[height=5.cm]{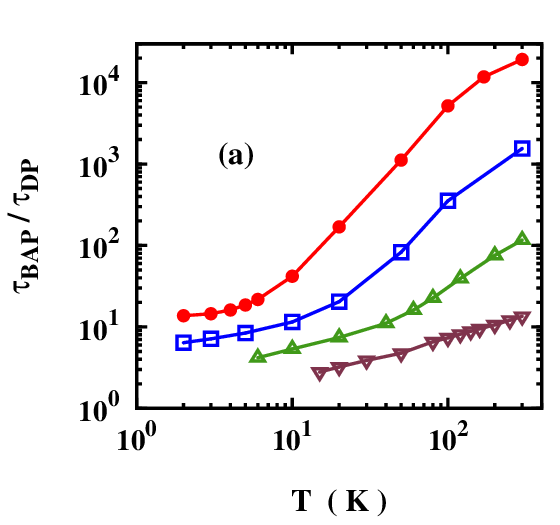}
\includegraphics[height=5.cm]{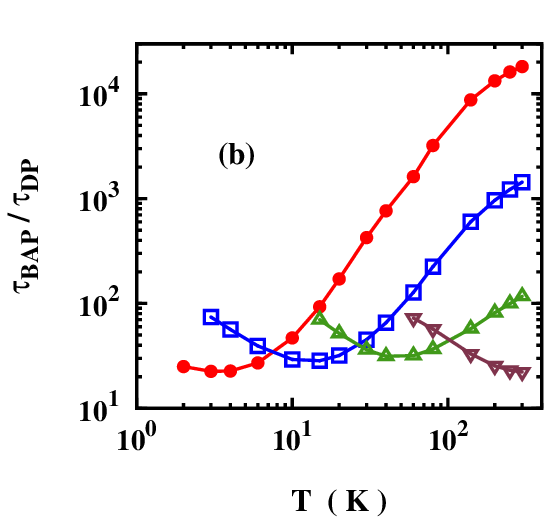}
\caption{(Color online) $p$-GaSb. Ratio of the SRT due to the BAP
  mechanism to that due to the DP mechanism as function of temperature
  for $n_h=3\times 10^{15}$~cm$^{-3}$ (curve with $\bullet$), 
  $3\times 10^{16}$~cm$^{-3}$ (curve with $\square$), 
  $3\times 10^{17}$~cm$^{-3}$ (curve with $\triangle$), 
  and $3\times 10^{18}$~cm$^{-3}$ (curve with $\triangledown$). 
  $n_i=n_h$. (a): $N_{\rm ex}=10^{14}$~cm$^{-3}$.  
(b): $N_{\rm ex}=0.1n_h$.}
\label{fig:T_pGaSb}
\end{figure}

\subsection{Other III-V semiconductors}

Although the above conclusions are obtained from GaAs, they also hold
for other III-V semiconductors. To demonstrate that, we also
investigate the problem in GaSb. GaSb is a narrow band gap III-V
semiconductors of which the values of $\Delta E_{\rm LT}$ and $\Delta
E_{\rm SR}$ can be found in literature.\cite{Aronov,bappara2} 
In Fig.~\ref{fig:T_pGaSb}, we plot the ratio of the SRT due
to the BAP mechanism to that due to the DP mechanism as function of
temperature for various hole densities. It is seen from the figures
that the features are similar to those in GaAs, whereas the ratio
 is much larger than that in GaAs under the same condition.
That is, the relative importance of the BAP mechanism in GaSb is
smaller than that in GaAs. This is because the SOC in GaSb is much
larger than that in GaAs while the longitudinal-transversal splitting
$\Delta E_{\rm LT}$ in GaSb is smaller than that in GaAs.

\section{Effects of electric field on spin relaxation in $n$-type
  III-V semiconductors}

\begin{figure}[htb]
\includegraphics[height=5.cm]{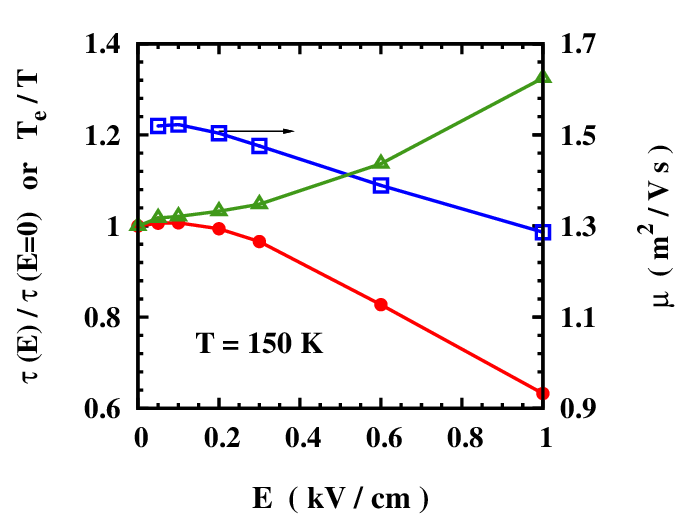}
\caption{(Color online) $n$-GaAs. Ratio of the SRT under electric
  field to the electric-field free one $\tau(E)/\tau(E=0)$ (curve
  with $\bullet$) and the ratio of the hot-electron temperature to the
  lattice temperature $T_e/T$ (curve with $\triangle$) as function of
  electric field for $n_i=n_{e}=2\times 10^{17}$~cm$^{-3}$ at
  $T=$150~K. The mobility is also plotted as curve with 
  $\square$ (note that the scale is on the right hand side of the
  frame).}
\label{fig:E1}
\end{figure}

In this section, we study the effects of electric field on spin
relaxation in $n$-type III-V semiconductors. Previous studies have
demonstrated that in quantum wells a relatively high in-plane electric
field can effectively manipulate the
SRT.\cite{hot-e,lowT,multi-band,multi-valley} The underlying physics 
is that the high electric field induces two 
effects: the drift of the electron ensemble which enhances the
inhomogeneous broadening (as electrons distribute on larger ${\bf
  k}$ states where the SOC is larger), as well as the hot-electron
effect which enhances the momentum scattering. The former tends to
suppress while the latter tends to enhance the SRT. Thus the SRT has
nonmonotonic electric field dependence: it first increases due to
the hot-electron effect then decreases due to the enhancement of
inhomogeneous broadening. In bulk semiconductors, the electric field
dependence of spin lifetime has not been investigated. In this section,
we present such a study. Using $n$-type GaAs as an example, we
demonstrate that the electric field dependence of spin lifetime can be
nonmonotonic (first increasing then decreasing) or monotonic
(decreasing) depending on the lattice temperature and the densities of
impurities and electrons. The underlying physics is analyzed. 
The study indicates that the spin lifetime can be effectively
controlled by electric field. 

\begin{figure}[htb]
\includegraphics[height=5.cm]{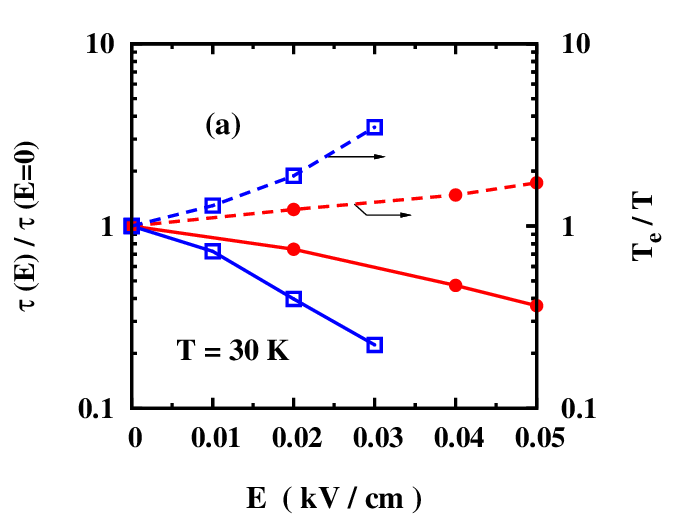}
\includegraphics[height=5.cm]{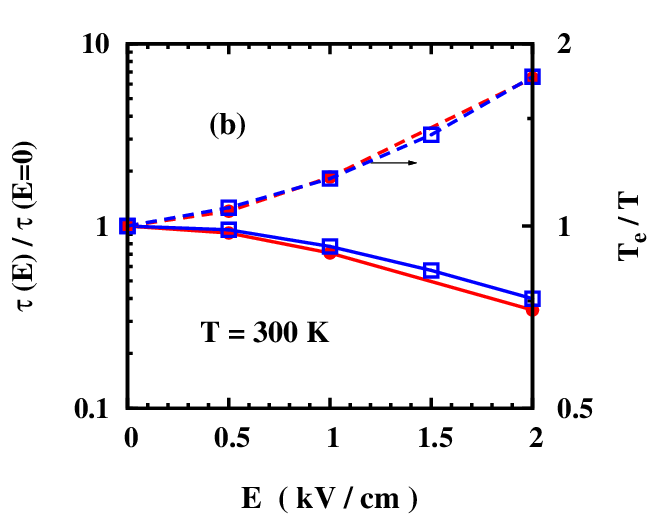}
\caption{(Color online) $n$-GaAs. Ratio of the SRT under electric
  field to the electric-field-free one $\tau(E)/\tau(E=0)$ (solid
  curves) and the ratio of the hot-electron temperature to the
  lattice temperature $T_e/T$ (dashed curves) as
  function of electric field for (a): $T=30$~K with
  $n_{e}=10^{16}$~cm$^{-3}$ (curve with $\bullet$: $n_i=n_{e}$; curve with
  $\square$: $n_i=0.05n_{e}$) and (b): $T=300$~K with $n_{e}=n_i$ (curve
  with $\bullet$: $n_{e}=10^{16}$~cm$^{-3}$; curve with $\square$:
  $n_{e}=2\times 10^{17}$~cm$^{-3}$).}
\label{fig:E2}
\end{figure}

In Fig.~\ref{fig:E1}, we plot the ratio of the SRT under electric field to
the electric-field-free one as function of electric field for
$n_{e}=2\times 10^{17}$~cm$^{-3}$ with $P=2~\%$ at
$T=150$~K and $n_i=n_{e}$. The electric field is chosen to be along the
$x$-axis and the initial spin polarization is along the $z$-axis. Due
to the cubic form of the SOC, the average of the spin-orbit field is
negligible even in the presence of finite center-of-mass drift
velocity, which is different from the case in quantum wells where the
linear ${\bf k}$-term gives a large effective magnetic field in the
presence of electric field thanks to the strong well
confinement.\cite{hot-e,multi-valley} It is seen that the ratio first increases a
little and than decreases rapidly with the electric field. At $E=1$~kV/cm, the ratio
drops to 0.6 which demonstrates that the electric field can effectively
affect the SRT. To understand these effects, we also plot the
hot-electron temperature in the figure. It is noted that the
electron temperature increases first slowly then ($E>0.3$~kV/cm)
rapidly with the electric field, indicating clearly the hot-electron
effect. For the 2DES, where the SOC is
dominated by the linear term, the hot-electron effect mainly leads to
the enhancement of scattering, whereas the enhancement of inhomogeneous
broadening due to the hot-electron effect is marginal.\cite{hot-e}
Differently, the hot-electron effect also has important effect on
inhomogeneous broadening in bulk system, as the SOC is cubic ${\bf k}$-dependent.
As both the drift effect and the hot-electron effect increase the
inhomogeneous broadening, the enhancement of the inhomogeneous
broadening is faster than the increase of momentum 
scattering. Consequently, the SRT is easier to decrease with the
electric field, which is different from the case of 2DES.
We speculate that the electric field dependence of the SRT in
Wurtzite semiconductors and strained III-V semiconductors
resembles that in the GaAs quantum wells, when the
SOC is dominated by the linear ${\bf k}$
term.\cite{wu-ZnO,wu-strain,Kato,ZnO} The mobility of the electron
system is also plotted in the figure. The variation of the mobility
indicates the nonlinear effects of the electric field in the kinetics
of electron system.

For lower and higher temperature cases, we plot the results in
Fig.~\ref{fig:E2}. The electron 
density is $n_{e}=10^{16}$~cm$^{-3}$ in Fig.~\ref{fig:E2}(a). It
is seen that the ratio for the case with low impurity density $n_i=0.05
n_{e}$ decreases faster than that for the case with $n_i=n_{e}$. This is
because both the hot-electron effect and the drift effect are more
pronounced in cleaner system,\cite{hot-e} which thus leads to a faster
decrease of the SRT due to the enhancement of inhomogeneous broadening.
In the background of hot-electron spin injection under high bias, our
results indicate that the manipulation of the SRT by the electric
field is more pronounced for samples with high mobility. For high
temperature case ($T=300$~K), we plot the SRT and hot-electron
temperature as function of electric field for two different electron
densities $n_{e}=10^{16}$~cm$^{-3}$ and $n_{e}=2\times 10^{17}$~cm$^{-3}$
with $n_i=n_{e}$ in Fig.~\ref{fig:E2}(b). It is seen
that both the SRT and the electron temperature differ marginally for
the two cases even though their impurity and electron densities differ by
20 times. This is because that at 300~K the electron--LO-phonon scattering is more
important than the electron-impurity scattering. Thus both the drift of the electron
system and the hot-electron effect is mainly determined by the electron--LO-phonon
scattering, and the variation of the SRT with electric field is
insensitive to impurity density. For electron density, as the electron
system is in the non-degenerate regime for both cases, the
electron density dependence is hence also weak.

\section{Conclusion}

In conclusion, we have applied an efficient scheme, the fully
microscopic KSBEs, to study the spin dynamics in bulk III-V
semiconductors, with all scatterings explicitly included. This approach
takes full account of the spin relaxation due to the DP, EY, and BAP
mechanisms in a fully microscopic fashion, and enables us to find
important results missing in the previous simplified approaches in the
literature. From the KSBE approach, we study the electron spin
relaxation in $n$-type, intrinsic, and $p$-type III-V
semiconductors. We also investigate the effects of electric field on
spin relaxation in $n$-type III-V semiconductors.

For $n$-type III-V semiconductors, differing from the previous conclusions, we
find that the spin relaxation due to the EY mechanism is less important
than that due to the DP mechanism even in narrow band-gap
semiconductors, such as InAs and InSb. We then focus on the spin
relaxation due to the DP mechanism. We find that the density
dependence of the SRT is nonmonotonic and we predict a peak which
appears in the metallic regime. This behavior is due to the different
density dependences of the inhomogeneous broadening and the momentum
scattering in the degenerate and non-degenerate
regimes. In the non-degenerate regime, as the electron distribution is
close to the Boltzmann distribution, the inhomogeneous broadening
changes little with the density but the electron-electron and
electron-impurity scatterings increase with the electron density. 
As a result, the SRT increases with the density. In the degenerate
regime, the inhomogeneous broadening increases with electron density, 
whereas the momentum scatterings decrease with it.
Consequently, the SRT decreases with the electron density in the
degenerate regime. The peak of the SRT is hence formed in the
crossover regime, where the corresponding Fermi temperature is close
to the lattice temperature, $T_{\rm F}\sim T$. Our results show that the
electron-electron scattering plays an important role in the spin
relaxation. We also study the density dependence for the case with
strain-induced SOC, where the density dependence of the inhomogeneous
broadening is different due to the linear ${\bf k}$-dependence of the
strain-induced SOC. However, the SRT still has a peak but at a
larger density compared to the strain-free case. We further
study the temperature dependence of the SRT. We find that the SRT
decreases monotonically with the temperature which is consistent
with experimental findings. After we
artificially lower the impurity density, we find a peak in the
SRT which is due to the different temperature
dependence of the electron-electron scattering in the degenerate and
non-degenerate regimes. This is consistent with the results in 2DES
where the peak in the SRT due to the
electron-electron scattering appears only when the impurity
density is low (e.g., $n_i=0.1n_{e}$).\cite{lowT} We also study the initial spin
polarization dependence of the SRT where the effect of the Coulomb HF
term is discussed. We find that the dependence is quite weak in bulk
system compared to that in the 2DES, which is again due to the large
impurity density $n_i\ge n_{e}$ in bulk system.

For intrinsic III-V semiconductors, we first compare the BAP mechanism and
the DP mechanism. We find that the BAP mechanism is far less efficient
than the DP mechanism. We further compare our results from the fully
microscopic KSBE approach with those from the approach widely
used in the literature. We find that the previous approach deviates in
the low temperature regime due to the pretermission of the Pauli
blocking. Also, the previous approach ignores the long-range
electron-hole exchange scattering which is shown to be dominant in
GaAs. We find that the electron-hole Coulomb scattering plays an important
role in spin relaxation. The Coulomb HF term is found to have
important effects on spin relaxation at low temperature and high 
photo-excitation density, as the impurity density is very low in intrinsic
semiconductors (we choose $n_i=0$). Due to the same reason, the peak in
the temperature dependence of the SRT due to the electron-electron scattering
also appears at small spin polarization. We further discuss the
photo-excitation density dependence of the SRT. We find that the SRT
first increases then decreases with the density which resembles the
case in $n$-type samples where the underlying physics is also similar.

For $p$-type III-V semiconductors, we first examine the relative importance
of the BAP mechanism. We find that the BAP mechanism 
dominates the spin relaxation in the low temperature regime only when the
photo-excitation density is low. However, when the photo-excitation
density is high, the BAP mechanism can be comparable
with the DP mechanism only in the moderate temperature regime
roughly around the Fermi temperature of electrons, whereas
for higher or lower temperature it is unimportant. The
photo-excitation density dependences of SRTs due to the BAP and DP
mechanisms are also discussed. We find that the relative importance of
the BAP mechanism decreases with photo-excitation density and
eventually becomes negligible at sufficiently high photo-excitation
density. For hole density dependence at small photo-excitation
density, we find that the spin relaxation due to the BAP mechanism
increases with hole density linearly in low hole density regime but
the increase becomes slower in high hole density regime
where the Pauli blocking of holes becomes important. 
Interestingly, the SRT due to the DP mechanism first increases, 
then decreases and again increases with hole density. The
underlying physics is that the momentum scattering (mainly from the
electron-impurity scattering) first increases with hole (impurity)
($n_i=n_h$) density, then decreases with hole density due to the
increase of the screening. However, at high hole density when the
hole system is degenerate, the screening increases slower with the hole
density and the momentum scattering again increases with the hole (impurity)
density. On the other hand, the inhomogeneous broadening does not
change with hole density as the electron density is solely determined by
the photo-excitation density. Consequently, the SRT due to the DP
mechanism first increases, then decreases and again increases with
the hole density. This behavior makes the ratio $\tau_{\rm
  BAP}/\tau_{\rm DP}$ first decreases rapidly, then slowly and again
rapidly with the hole density. The BAP mechanism is more
  important than the DP one for high hole density.
The relative importance of the BAP mechanism in GaSb is found to
be much less than that in GaAs due to both the weaker electron-hole
exchange interaction and the larger SOC in GaSb.

Finally, we study the effect of electric field on the spin relaxation
in $n$-type GaAs. We find that the SRT can be largely affected by the
electric field. The underlying physics is that the electric field
induces two effects: the center-of-mass drift which enhances the
inhomogeneous broadening and the hot-electron effect which increases
both the momentum scattering and the inhomogeneous broadening. The
electric field dependence of SRT thus can be nonmonotonic: it first
increases due to the increase of scattering then decreases due to the
enhancement of inhomogeneous broadening. However, we find that
differing from the 2DES, the SRT is easier to decrease with the
electric field. This is because that the inhomogeneous broadening
increases faster when the SOC is cubic compared to the 2DES where the
SOC is dominated by linear ${\bf k}$ term. We expect that the electric
field dependence of the SRT resembles the 2DES for Wurtzite
semiconductors or strained semiconductors, where the SOC can be
dominated by the linear ${\bf k}$ term. We also find that the effect
of the electric field becomes more significant for low impurity density
samples at low temperature as both the drift effect and the
hot-electron effect are more pronounced. However, at room temperature,
the effect of the electric field is insensitive to the impurity
density as the electron--LO-phonon scattering is more important than
the electron-impurity scattering. The electron density dependence is
also weak as long as the system is in the non-degenerate regime.

Note added: After we submitted this manuscript, the peak we predicted
in the doping density dependence of the SRT of $n$-GaAs in the
metallic regime has been realized experimentally in a subsequent
paper.\cite{nGaAs-peak} Also an independent theoretical calculation
from the KSBE approach well reproduced the peak in the same paper.\cite{nGaAs-peak}

\begin{acknowledgments}
  This work was supported by the Natural Science Foundation of China
  under Grant No.~10725417, the National Basic Research
  Program of China under Grant No.~2006CB922005, and the
  Knowledge Innovation Project of Chinese Academy of Sciences. One of
  the authors (J.H.J.) would like to thank M. Q. Weng and K. Shen for
  helpful discussions.
\end{acknowledgments}

\begin{appendix}

\section{Numerical scheme}

Our numerical scheme is based on the discretization of the 
${\bf k}$-space similar to that in Ref.~\onlinecite{hot-e}. 
Here we extend it to the three dimensional case. Our technique greatly
reduces the calculation complexity and makes the quantitatively
accurate calculation possible.

The ${\bf k}$-space is divided
into $N\times M\times L$ control regions where the ${\bf k}$-grid points
are chosen to be ${\bf k}_{n,m,l}=\sqrt{2 m_c E_{n}} (\sin \theta_m
\cos \phi_l, \sin \theta_m\sin \phi_l, \cos \theta_m )$. To facilitate
the evaluation of the $\delta$-functions in the scattering terms, we set
$E_{n}=(n+1/2)\Delta E$, where the energy span in
each control region is $\Delta E = \omega_{\rm LO}/n_{\rm LO}$ with $n_{\rm LO}$
being an integer number and $\omega_{\rm LO}$ denoting the LO-phonon
frequency. The electron-impurity and electron-phonon scatterings are
then solved easily since the $\delta$-functions can be integrated out
directly.

For the electron-electron Coulomb, electron-hole Coulomb and electron-hole
exchange scatterings, the situation is much more complex. The
electron-electron Coulomb scattering term [Eq.~(\ref{scat_ee})] can be
rewritten as,
\begin{eqnarray}
 \left. \partial_t \hat{\rho}_{\bf k}\right|_{\rm ee} &=& \sum_{{\bf k}^{\prime}}
 \frac{-\pi}{(2\pi)^3} V_{{\bf k}-{\bf k}^{\prime}}
 \big[\hat{\Lambda}_{{\bf k},{\bf k}^{\prime}}\hat{\rho}_{{\bf
 k}^{\prime}}^{>}\hat{\Lambda}_{{\bf k}^{\prime},{\bf
 k}}\hat{\rho}_{\bf k}^{<}H({\bf k},{\bf k}^{\prime}) \nonumber\\
 &&\mbox{} -\hat{\Lambda}_{{\bf k},{\bf k}^{\prime}}
\hat{\rho}_{{\bf k}^{\prime}}^{<}\hat{\Lambda}_{{\bf k}^{\prime},{\bf
 k}}\hat{\rho}_{\bf k}^{>}H({\bf k}^{\prime},{\bf k})\big] + {\rm H.c.},
\end{eqnarray}
where 
\begin{eqnarray}
&&\hspace{-1.1cm}  H({\bf k},{\bf k}^{\prime}) = (2\pi)^3 \sum_{{\bf k}^{\prime\prime}}
  \delta(\varepsilon_{{\bf k}^{\prime\prime}}-\varepsilon_{{\bf k}^{\prime\prime}-{\bf k}+{\bf k}^{\prime}}+\varepsilon_{{\bf k}^{\prime}}-\varepsilon_{\bf k}) \nonumber\\ 
&&\hspace{-0.3cm} \times \mbox{} {\rm Tr}\left(\hat{\Lambda}_{{\bf
  k}^{\prime\prime},{\bf k}^{\prime\prime}-{\bf k}+{\bf k}^{\prime}}\hat{\rho}_{{\bf k}^{\prime\prime}-{\bf k}+{\bf k}^{\prime}}^{<}\hat{\Lambda}_{{\bf k}^{\prime\prime}-{\bf k}+{\bf k}^{\prime},{\bf k}^{\prime\prime}} \hat{\rho}_{{\bf k}^{\prime\prime}}^{>}
  \right).
\end{eqnarray}
Substituting ${\bf q}={\bf k}-{\bf k}^{\prime}$ and $\omega=2 m_c
(\varepsilon_{\bf k}-\varepsilon_{{\bf k}^{\prime}})$, one has
\begin{eqnarray}
\hspace{-0.4cm}  H({\bf q},\omega) &=& \int \!\!\ d{{\bf k}^{\prime\prime}} \delta(\varepsilon_{{\bf k}^{\prime\prime}}-\varepsilon_{{\bf k}^{\prime\prime}-{\bf q}} - \omega/2m_c
  ) \nonumber\\  
&& \times \mbox{} {\rm Tr}\left(\hat{\Lambda}_{{\bf
  k}^{\prime\prime},{\bf k}^{\prime\prime}-{\bf q}}\hat{\rho}_{{\bf k}^{\prime\prime}-{\bf q}}^{<}\hat{\Lambda}_{{\bf k}^{\prime\prime}-{\bf q},{\bf k}^{\prime\prime}} \hat{\rho}_{{\bf k}^{\prime\prime}}^{>}
  \right).
\end{eqnarray}
Now the $\delta$ function can be simplified as
\begin{equation}
  \delta(\varepsilon_{{\bf k}^{\prime\prime}}-\varepsilon_{{\bf k}^{\prime\prime}-{\bf q}} - \omega/2m_c
  )  = \frac{m_c}{ {k}^{\prime\prime} q} \delta( \cos\hat{\theta}-\cos\hat{\theta}_0),
\end{equation}
where $\hat{\theta}$ is the angle between ${\bf k}^{\prime\prime}$ and
${\bf q}$, and
$\cos\hat{\theta}_0=(q^2+\omega)/(2{k}^{\prime\prime}q)$. To evaluate
the $\delta$ function, it is helpful to rotate to the new coordinate system
with ${\bf q}$ being along the $z$-axis. In this coordinate system,
$\hat{\theta}=\theta$ and the $\delta$ function can be evaluated readily.
The result is
\begin{eqnarray}
&&\hspace{-1.1cm}  H({\bf q},\omega) = \frac{m_c^2}{q} \int \!\!\
  d{\varepsilon_{{\bf k}^{\prime\prime}}} d{\phi^{\prime\prime}} 
\nonumber\\ &&\hspace{0.4cm}  \times
  {\rm Tr} \Big(\hat{\Lambda}_{{\bf
  k}^{\prime\prime},{\bf k}^{\prime\prime}-{\bf q}}\hat{\rho}_{{\bf k}^{\prime\prime}-{\bf q}}^{<}\hat{\Lambda}_{{\bf k}^{\prime\prime}-{\bf q},{\bf k}^{\prime\prime}} \hat{\rho}_{{\bf k}^{\prime\prime}}^{>}
  \Big) \bigg|_{\theta=\theta_0}^{({\rm new})},
\end{eqnarray}
with $\theta_0=\arccos [(q^2+\omega)/2{k}^{\prime\prime}q] $. Note that the
integration over $\varepsilon_{{\bf k}^{\prime\prime}}$ is restrained by the condition
$\varepsilon_{{\bf k}^{\prime\prime}}\ge [(q^2+\omega)/2q]^2/2m_c$ according to
$|\cos\theta_0|^2\le 1$.
Now the electron-electron Coulomb scattering is easily integrated
out. Note that the calculation of $H({\bf q},\omega)$ can be done
before the calculation of the electron-electron Coulomb scattering
terms, which thus reduces the whole calculation complexity from
$O(N^2M^3L^3)$ to $O(N^2M^2L^2)$. This method, first developed by
Cheng in two-dimensional system,\cite{Cheng} greatly reduces the calculation complexity.

For the electron-hole Coulomb scattering, the idea is similar, but the
technique is more complex. Denoting $x_m=m_c/m^{\ast}_m$, the
$\delta$-function in Eq.~(\ref{scat_eh}) can be written as
\begin{eqnarray}
&&\hspace{-1.cm}  \delta(\varepsilon^h_{{\bf k}^{\prime\prime}m}-\varepsilon^h_{{\bf k}^{\prime\prime}-{\bf q} m^{\prime}} - \omega/2m_c
  ) = 2 m_c \nonumber \\
&& \hspace{-0.95cm} \times \delta\left( \left(x_m-x_{m^{\prime}}\right){{k}^{\prime\prime}}^2 - x_{m^{\prime}} q^2 + 2
  x_{m^{\prime}} {k}^{\prime\prime}q \cos \hat{\theta} - \omega \right).
\end{eqnarray}
The $\delta$-function is then integrated out similarly,
\begin{eqnarray}
  H_{\rm eh}({\bf q},\omega) &=& \sum_{m,m^{\prime}}\frac{m_c^2}{q x_{m} x_{m^{\prime}}} 
\int \!\!\ d{\varepsilon^h_{{\bf k}^{\prime\prime}}} d{\phi^{\prime\prime}}  |{\cal T}^{{\bf
    k}^{\prime\prime}m}_{{\bf k}^{\prime\prime}-{\bf q} m^{\prime}}|^2 \nonumber \\
&& \times \mbox{} 
      f^{h}_{{\bf k}^{\prime\prime}-{\bf q} m^{\prime}}       
      \left(1 - f^{h}_{{\bf k}^{\prime\prime}m}\right)
      \bigg|_{\theta=\theta_0}^{({\rm new})}. 
\end{eqnarray}
The integration over $\varepsilon^h_{{\bf k}^{\prime\prime}}$ is restrained by the condition
$|\cos \theta_0|^2\le 1$, where 
\begin{equation}
  \cos \theta_0 = \frac{1}{2 {k}^{\prime\prime} q} \left(
  \frac{\omega}{x_{m^{\prime}}} + q^2 - \frac{ x_{m} -
  x_{m^{\prime}} }{x_{m^{\prime}}} {{k}^{\prime\prime}}^2\right).
\end{equation}
The restriction can be simplified as
\begin{equation}
 b^2 {{k}^{\prime\prime}}^2 - (1+2ab) {{k}^{\prime\prime}} + a^2\le 0 \quad {\rm and} \quad {k}^{\prime\prime}\ge 0,
\end{equation}
where $a=(\frac{\omega}{x_{m^{\prime}}}+q^2)/(2q)$ and
$b=(x_{m}-x_{m^{\prime}})/(2x_{m^{\prime}}q)$. The above inequality can be worked out
readily for given $(q,\omega,m,m^{\prime})$, and the restriction condition for
$\varepsilon^h_{{\bf k}^{\prime\prime}}= x_{m}
{{k}^{\prime\prime}}^2/(2m_c)$ is then obtained.

The electron-hole exchange scattering is solved similarly by substituting the
matrix element $|{\cal T}^{{\bf k}^{\prime\prime}m}_{{\bf
    k}^{\prime\prime}-{\bf q}m^{\prime}}|^2$ with
$|{\cal J}^{(\pm)\ {\bf
    k}^{\prime\prime}m}_{{\bf k}^{\prime\prime}-{\bf 
    q}m^{\prime}}|^2$. Finally, the drift term is
solved with similar method of that in Ref.~\onlinecite{hot-e}. The
differential equations are solved by the fourth-order Runge-Kutta method.

The computation is carried out in a parallel manner by using
OpenMP. For a typical calculation with the partition of
$40\times8\times16$ grid points in the ${\bf k}$-space, it takes about
ten hours to evolute 50~ps on a quad-core AMD Phenom 9750.

\section{Comparison with experiment}

\begin{figure}[htb]
\includegraphics[height=5.cm]{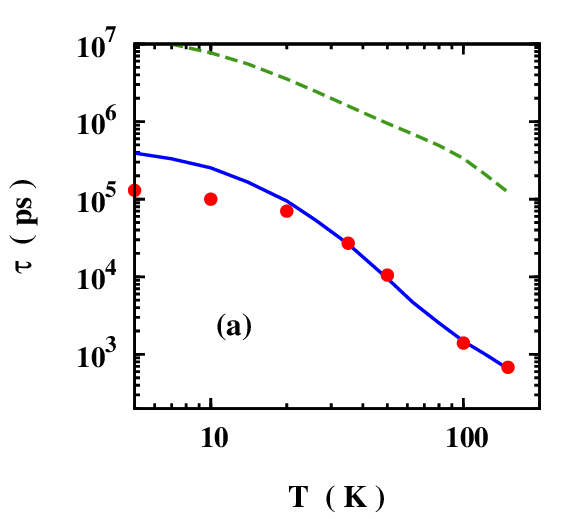}
\includegraphics[height=5.cm]{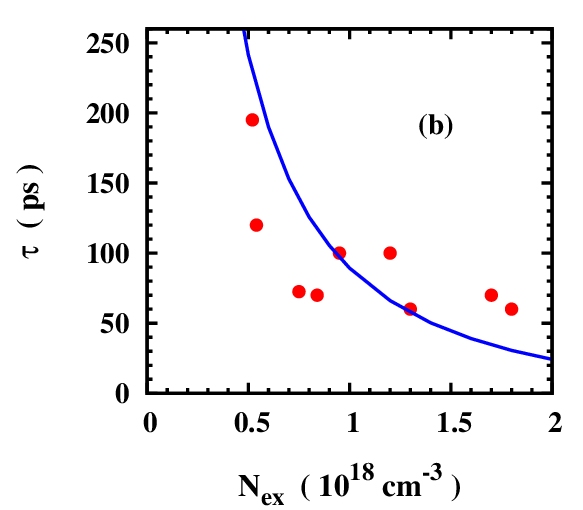}
\includegraphics[height=5.2cm]{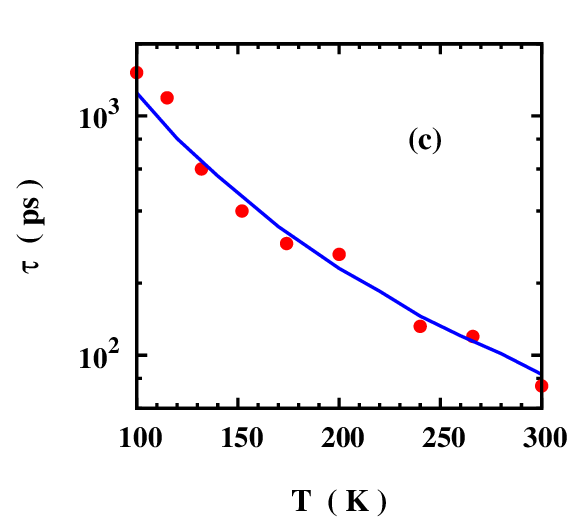}
\caption{(Color online) (a) $n$-GaAs. SRTs $\tau$ from the
  experiment in Ref.~\onlinecite{Awsch98} ($\bullet$) and
  from the calculation via the KSBE approach with only the DP
  mechanism (solid curve) as well as that with only the EY
  mechanism (dashed curve). $n_{e}=10^{16}$~cm$^{-3}$, $n_i=n_e$ and
  $N_{\rm ex}=10^{14}$~cm$^{-3}$. $\gamma_{\rm
  D}=8.2$~eV$\cdot$\AA$^{3}$. (b) $p$-GaAs. SRTs $\tau$ from the 
  experiment in Ref.~\onlinecite{Nex-decr} ($\bullet$) and
  from the calculation via the KSBE approach (solid
  curve). $n_h=6\times10^{16}$~cm$^{-3}$, $n_i=n_h$ and
  $T=100$~K. $\gamma_{\rm D}=8.2$~eV$\cdot$\AA$^{3}$. (c)
  $p$-GaAs. SRTs $\tau$ from the experiment in
  Ref.~\onlinecite{Zerrouati} ($\bullet$) and from the calculation via
  the KSBE approach (solid curve). $n_h=1.6\times10^{16}$~cm$^{-3}$,
  $n_i=n_h$ and $N_{\rm ex}=10^{14}$~cm$^{-3}$. $\gamma_{\rm
  D}=10$~eV$\cdot$\AA$^{3}$.} 
\label{fig:exp}
\end{figure}

We compare the calculation from the fully microscopic KSBE approach
with the experimental results in Refs.~\onlinecite{Awsch98} and
\onlinecite{Nex-decr}. These experiments were carried out in $n$-
and $p$-type GaAs respectively. In Fig.~\ref{fig:exp}(a), we plot the SRT
as function of temperature calculated from the KSBEs, together with the
experimental data in Ref.~\onlinecite{Awsch98} for $n$-GaAs with
$n_{e}=10^{16}$~cm$^{-3}$, $n_i=n_e$ and $N_{\rm
  ex}=10^{14}$~cm$^{-3}$. It is seen that the 
calculation agrees well with the experimental results in $n$-GaAs for
$T\gtrsim 20$~K. The deviation in the lower temperature regime is due
to the rising of the localization of
electrons.\cite{localize,opt-nmr,localize2,localizeReview} The SRT due to the EY
mechanism is also plotted in the figure, which is much larger than the
experimental data, indicating the irrelevance of the EY mechanism.
The calculation gives a fit of the SOC parameter as
$\gamma_{\rm D}=8.2$~eV$\cdot$\AA$^{3}$ which is different from the value 
$\gamma_{\rm D}=23.9$~eV$\cdot$\AA$^{3}$ calculated from the tight-binding
or ${\bf k}\cdot{\bf p}$ parametric theories.\cite{wu-soc,soc35}
However, the value is still in the reasonable range of $\gamma_{\rm D}$
[for lists of $\gamma_{\rm D}$ calculated and measured via various methods,
see Ref.~\onlinecite{listsoc}]. Our fitting, though with only one fitting
parameter $\gamma_{\rm D}$, agrees well with the experimental data in
almost the whole temperature range, which is much better than the fittings
with the same experimental data in
Refs.~\onlinecite{GaAs-cal1}-\onlinecite{GaAs-cal3}. In 
Fig.~\ref{fig:exp}(b), we plot the photo-excitation density
  dependence of the SRTs from our calculation and from the experiment in
Ref.~\onlinecite{Nex-decr} for $p$-GaAs with
  $n_h=6\times10^{16}$~cm$^{-3}$, $n_i=n_h$ and $T=100$~K. It is seen
 that with the same $\gamma_{\rm D}$, the calculation again agrees
 well with the experimental data. The SRT due to the BAP mechanism is
  about 20~times larger, which is 
  consistent with our conclusion that the BAP mechanism is unimportant
  at high photo-excitation density. In Fig.~\ref{fig:exp}(c), we plot
  the temperature dependence of the SRTs from our calculation and from
  the experiment in Ref.~\onlinecite{Zerrouati} for $p$-GaAs with
  $n_h=1.6\times10^{16}$~cm$^{-3}$, $n_i=n_h$ and $N_{\rm
    ex}=10^{14}$~cm$^{-3}$. The best fitting gives a slightly larger
  $\gamma_{\rm D}=10$~eV$\cdot$\AA$^{3}$. The SRT due to the BAP
  mechanism is $\sim$100~times larger than that due to the DP
  mechanism. This is consistent with our conclusion that the BAP
  mechanism is unimportant at high temperature for low doping density.

Throughout the paper, the SRT in GaAs due to the DP mechanism is
calculated with $\gamma_{\rm D}=23.9$~eV$\cdot$\AA$^{3}$  unless in
Sec.~V where we use $\gamma_{\rm D}=8.2$~eV$\cdot$\AA$^{3}$ in order to
avoid possible exaggeration of the DP mechanism by using a ``larger''
SOC parameter. However, the SRT due to the DP mechanism is proportional to
$\gamma_{\rm D}^2$ in motional narrowing regime, thus the results presented
in the paper can be easily converted to the results for another
$\gamma_{\rm D}$. The ratio for the SRTs in the two cases is 8.5.

\section{Role of screening on DP spin relaxation in $p$-type GaAs}

\begin{figure}[htb]
\includegraphics[height=5.cm]{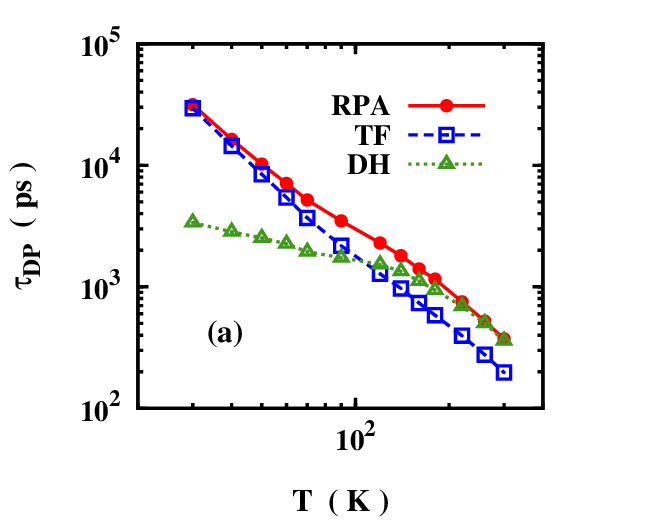}
\includegraphics[height=5.cm]{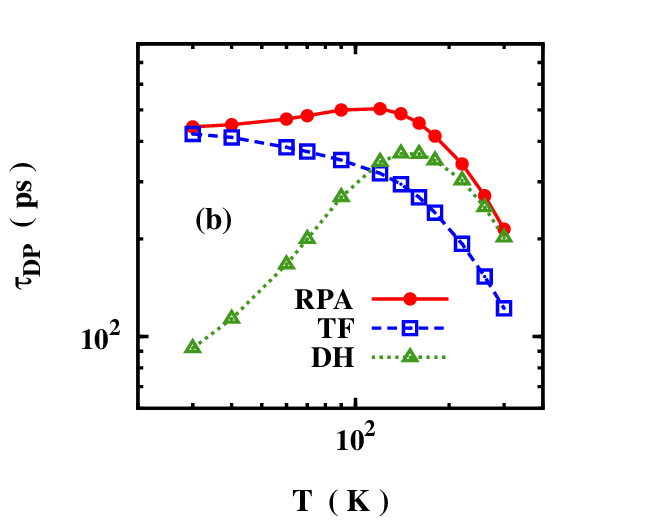}
\caption{(Color online) $p$-GaAs. The SRTs due to the DP mechanism
  with the TF (curve with $\square$), DH (curve with $\triangle$), and
  the RPA (curve with $\bullet$) screenings for excitation density (a)
  $N_{\rm ex}=10^{14}$~cm$^{-3}$ and (b)
  $N_{\rm ex}=3\times10^{17}$~cm$^{-3}$. Hole density $n_h=3\times
  10^{18}$~cm$^{-3}$ and $n_i=n_h$.}
\label{fig:pscreen}
\end{figure}

We study the effect of screening on DP spin relaxation in $p$-type
GaAs. We focus on the temperature dependence of the SRT which
corresponds to the discussions on the results in
Figs.~\ref{fig:T_p} and \ref{fig:T_p2}. As in Sec.~V, we
discuss both the low and high excitation density cases. We first study
the low excitation case. To elucidate the role of screening on the DP spin
relaxation, we plot the SRT due to the DP mechanism with the
TF screening\cite{koch} [which applies in the degenerate (low temperature)
  regime], the DH screening\cite{koch} [which applies in the
  non-degenerate (high temperature) regime], and the RPA
screening\cite{koch} (which applies in the whole temperature regime)
in Fig.~\ref{fig:pscreen}(a). The crossover 
from the DH screening to the TF one is clearly seen in the
figure. Also, compared with the case of temperature-independent
screening (i.e., the TF screening), one can see that the increase of the
SRT with decreasing temperature is indeed slowed down around $T_{\rm F}^{h}$
(where the screening effect is prominent) due to the increase of
screening.

For high excitation case, we find a peak in the temperature dependence
of the SRT due to the DP mechanism. This peak is roughly around $T_{\rm F}$. To
explore the underlying physics, we also plot the SRT due to the DP
mechanism with the TF and DH screenings. It is seen that with the TF
screening the peak disappears, while with the DH screening the peak
remains. This elucidates that the appearance of the peak is due
to the increase of screening at low temperature. What is different for
the high excitation case is that the electron Fermi temperature is
higher than the hole Fermi temperature. Thus at low temperature when
the electron system is in degenerate regime (hence inhomogeneous
broadening changes slowly) and the hole system has not yet entered into the
degenerate regime (hence the increase of screening with decreasing
temperature is prominent), the SRT due to the DP mechanism $\tau_{\rm DP}$ decreases
with decreasing temperature. On the other hand, at high temperature, 
$\tau_{\rm DP}$ increases with decreasing temperature. Consequently,
the peak is formed. It should be noted that this peak is different
from the peak found in intrinsic semiconductors in Sec.~IV where the
peak is due to the nonmonotonic temperature dependence of the
electron-electron Coulomb scattering, whereas the main scattering
mechanism here is the electron-impurity scattering.

\end{appendix}


\begin{thebibliography}{0}
\bibitem{note9} The spin-orbit parameter $\gamma_{\rm D}$ in GaAs fitted from
  experiments is 2.9~times smaller than that in Table I. Even for such a
  $\gamma_{\rm D}$, the spin relaxation due to the DP mechanism is still much
  more efficient ($\gtrsim 20$~times) than that due to the EY
  mechanism [see also Appendix~B].
\bibitem{note7} Note that the SOC parameter we used in this section
  ($\gamma_{\rm D}=23.9$~eV$\cdot$\AA$^{3}$) is larger than the value
  ($\gamma_{\rm D}=8.2$~eV$\cdot$\AA$^{3}$) fitted from experimental data
  (see Appendix~B). However, as the DP spin relaxation is more than
  two orders of magnitude larger than the BAP spin relaxation, the
  choice of the $\gamma_{\rm D}$ does not change the conclusion that the BAP
  mechanism is unimportant.
\bibitem{note6} Further calculation indicates that the results from
  Eq.~(\ref{BAPelastapp1}) deviate more from the calculation via the
  KSBEs at small spin polarization as the Pauli blocking of spin-flip
  scattering is more pronounced when both spin bands are nearly
  equally populated.
\bibitem{HFtaup} According to Eq.~(\ref{HFeq}), the
 effect of the Coulomb HF term can be estimated by $\tau_s(P) =
 \tau_s(P=0)\left[1+(V_{q_{0}} n_{e} P \tau_p^{\ast})^2 \right]$ where
 $V_{q_{0}}$ describes the average Coulomb interaction with
 $q_{0}$, $n_{e}$, $P$ being the average momentum, the electron density,
 and the initial spin polarization, respectively. Thus the effect of the Coulomb HF
 term is more pronounced for larger $n_{e}$, $P$ and weaker scattering
 (i.e., larger $\tau_p^{\ast}$). Moreover, increase of the temperature would
 reduce the effect as $q_{0}$ becomes larger and $V_{q_{0}}$ becomes
 smaller.
\bibitem{note3} The EY mechanism is checked to be unimportant in the
  parameter region we studied for both $p$-GaAs and $p$-GaSb.
\bibitem{note8} The initial spin polarization dependence of the SRT is
  weak due to the fact that the impurity density is larger than the
  electron density according to the analysis in Sec.~III~B3.
\end{thebibliography}
\end{document}